\documentclass{article}

\usepackage{arxiv}
\usepackage[utf8]{inputenc} 
\usepackage[T1]{fontenc}    
\usepackage{hyperref}       
\usepackage{booktabs}       
\usepackage{amsfonts}       
\usepackage{graphicx}
\usepackage{subfigure}
\usepackage{authblk}
\usepackage{blindtext}

\usepackage{pdfpages}
\usepackage{subfigure}

\usepackage{booktabs}

\begin{document}

\title{Identification of Key Companies for International Profit Shifting in the Global Ownership Network}

\makeatletter

\author{Tembo Nakamoto \textsuperscript{1 a}, Abhijit Chakraborty \textsuperscript{2 b}, Yuichi Ikeda \textsuperscript{3 a}}

\renewcommand\@date{{%
  \vspace{-\baselineskip}%

 \textsuperscript{1}{nakamoto.tembo.75w@st.kyoto-u.ac.jp}\par
 \textsuperscript{2}{abhiphyiitg@gmail.com}\par
 \textsuperscript{3}{ikeda.yuichi.2w@kyoto-u.ac.jp}

 \bigskip

  \textsuperscript{a}{\normalsize Graduate school of Advanced Integrated Studies in Human Survivability, Kyoto University, 1, Yoshida-Nakaadachi-cho, Sakyo-ku, 6068306, Kyoto, Japan}\par
  \textsuperscript{b}\normalsize Graduate School of Simulation Studies, University of Hyogo, 7-1-28, Minatojima-minamimachi, Chuo-ku, 6500047, Kobe, Japan

  \bigskip

 \date{}
}}
\makeatother

\maketitle

\begin{abstract}
In the global economy, the intermediate companies owned by multinational corporations are becoming an important policy issue as they are likely to cause international profit shifting and diversion of foreign direct investments. The purpose of this analysis is to call the intermediate companies with high risk of international profit shifting as key firms and to identifying and clarify them. For this aim, we propose a model that focuses on each affiliate's position on the ownership structure of each multinational corporation. Based on the information contained in the Orbis database, we constructed the Global Ownership Network, reflecting the relationship that can give significant influence to a firm, and analyzed for large multinational corporations listed in Fortune Global 500. In this analysis, first, we confirmed the validity of this model by identifying affiliates playing an important role in international tax avoidance at a certain degree. Secondly, intermediate companies are mainly found in the Netherlands and the United Kingdom, etc., and tended to be located in jurisdictions favorable to treaty shopping. And it was found that such key firms are concentrated on the IN component of the bow-tie structure that the giant weakly connected component of the Global Ownership Network consist of. Therefore, it clarifies that the key firms are geographically located in specific jurisdictions, and concentrates on specific components in the Global Ownership Network. The location of key firms are related with the ease of treaty shopping, and there is a difference in the jurisdiction where key firms are located depending on the location of the multinational corporations.
\end{abstract}


\section*{Introduction}
Multinational corporations (MNCs) and their affiliates in various jurisdictions have become important players in the globalized world. However, they have been recognized as important players in policy issues and their ownership structures have become more complicated (UNCTAD 2016). The presence of intermediate companies is the major factors contributing to such complex structures. MNCs usually operate through their local affiliates and indirectly through intermediate companies. Various factors influence their decision for using such intermediate companies for their business operations. Improving management efficiency and avoiding legal risks are some of the reasons: however, the biggest reason is to minimize withholding tax liability (Eicke 2009).\par
Basically, each jurisdiction imposes withholding tax on the profits relocated to other jurisdictions; however, if two jurisdictions conclude a tax treaty and agree to offer a withholding tax reduction to companies in each other jurisdiction, companies located in those two jurisdictions can enjoy these tax benefits (Arnold 2016). Therefore, only companies located in the jurisdictions of contracting a tax treaty are provided with the relief of withholding tax. Companies located in other jurisdictions are not qualified to enjoy this withholding tax reduction because their jurisdictions do not conclude a tax treaty. To meet this withholding tax reduction requirements, MNCs often establish intermediate companies in the jurisdictions that offer these reliefs despite conducting no reasonable business purpose in these jurisdictions (Vega Borrego 2006). This technique is called treaty shopping using intermediate companies to minimize withholding tax liability. Treaty shopping is one of the schemes of international profit shifting and a scheme against the fairness of tax liability (Avi-Yonah and Panayi 2010). Many MNCs avoid tax liability by establishing intermediate companies and making ownership structures complicated (Maine and Nguyen 2017).\par
Moreover, intermediate companies are indispensable elements for taxation purpose and for the cross-border investment. MNCs also create foreign direct investment (FDI) through their affiliates or intermediate companies (Weyzig 2013). Although investment is made through a third-party country, statistics on FDI are in bilateral units, which cannot reveal the actual amount of FDI. The Organisation for Economic Co-operation and Development recommends that member countries should separate detoured FDI from the actual amount of FDI (Borga 2015) because the detours are in a large global economy and this is difficult to be ignored.The reason is based on the MNCs' indirect ownership of their intermediate companies.\par
Thus, it is important to understand how the tax policy of each jurisdiction influences the flow of capital inside MNCs for tackling international profit shifting. In addition, it is essential to grasp detoured FDI made through intermediate companies to understand the realities of the current global economy. On the basis of this background, we analyze the activities of intermediate companies.\par
Since the 1990s, network science has made remarkable progress (Newman 2003; Holme and Saram$\ddot{a}$ki 2012 et al.) and has presented a new way of looking at the entire society, revealing that the relationship between the elements or actors determines the behavior of each element or actor. In recent years, the development of economic data has reached an exponential growth, and the ownership of the network that represents a special type of economic networks is one of such examples. The ownership network has been analyzed from various aspects such as the topology of shareholding relationships in the Italian and the United States stock markets (Battiston 2004); the control of ownership among European companies (Glattfelder and Battiston 2009) in geographical space (Vitali and Battiston 2011). The architecture and structure of the ownership network is an effective way to assess the complexity of the ownership such as transnational corporations (Vitali et al. 2011; Vitali and Battiston 2014). Especially, understanding the complexity of the ownership hierarchies of large bank holding companies assist in reaching a resolution is required when a company goes bankrupt (Flood et al. 2017). The ownership network also is applied to the context of international taxation and analysis of offshore financial centers (Gracia-Bernardo et al. 2017). Comprehending connectivities in the real world helps us to understand the economic growth and incomes more clearly (Gould et al. 2018).\par
Many studies on international taxation focus on macroeconomic data. However, such analyses have limitations because macroeconomic data contain many noises and biases. Since FDI is related to bilateral units, it is difficult to capture the diversion of FDI perpetuated by MNCs for treaty shopping. In recent years, many studies have used microdata because of the exponential growth of data. However, most of the studies use ownership information to determine whether a company is an affiliate of an MNC, and these studies do not consider ownership relations of affiliates. Thus, the information is not enough to understand the diversion of FDI. Although a few studies considered the ownership relations among affiliates, however, they only analyzed MNCs located in specific countries (Collins and Shackelford 1998; Mintz and Weichenrieder 2010). Their analysis was not on global trends. To the best of our knowledge, only one study adopts the network science approach in conducting a global analysis of ownership relations (Garcia-Bernardo et al. 2017). However, their approach is designed on the trend of jurisdictions not paying attention to each affiliate.\par
The purpose of this paper is to investigate the complicated ownership structure of MNCs from the viewpoint of network science, considering their attitude toward tax and international profit shifting. We refer to intermediate companies as key companies involved in a high-risk international profit shifting. Thus, we propose a model that hierarchically identifies these companies as affiliates of MNCs. The feature of this model is to consider ownership relations among affiliates and the position of affiliates in the ownership structure of each MNC. We newly introduce the concept called layer to identify the key companies that are at risk in international profit shifting.\par
The structure of this paper is as follows: In the Dataset section, we mention the databases that are used to build the Global Ownership Network (GON). In the Model section, we propose and explain the model to identify the key companies that are at high risk in international profit shifting. In the Results section, we describe the structure of the GON and the locations of key companies described in our model. In the Discussion section, we discuss the results in the context of international taxation. In the Conclusion section, we explain our analysis briefly and outline important results and their implications.

\section*{Dataset}
We construct the GON based on the ownership information in the Orbis 2015 database. Generally, the larger MNCs face a high risk in international profit shifting. We analyze MNCs listed in the Fortune Global 500.

\subsection*{Orbis Database}
In this analysis, we use the Orbis database (Bureau van Dijk 2015), which comprises financial information of 59,581,452 companies and individuals located across 215 jurisdictions. Information on the balance sheet, profit and loss accounts, the standard industrial classification (SIC), geographical position, and  the ownership structure are provided.
\par
The database contains information of each company as it is reported to the Ministry of Commerce of each jurisdiction. The recording rates of company's information vary greatly from jurisdiction to jurisdiction and this depends on the size of the companies. Small companies are less likely to be included because of their lack of accuracy (Kalemi-Ozcan et al. 2015). Moreover, companies located in low-income jurisdictions are less likely to be included (Cobham \& Loretz 2014). Lack of financial data in certain no-corporate-tax jurisdictions is a serious issue, but this can be mitigated the ownership relations, although all ownership relationships still are not included in the database. Moreover, it is difficult to evaluate the magnitude and importance of the missing ownership relationships due to a general lack of data on actual ownership relations. However, an important number of ownership relations in no-corporate tax jurisdictions are identified among MNCs listed in the Fortune Global 500 (OECD 2015). Although the database suffers from potential biases, it has been widely used for tax empirical analyses because it is considered as the most comprehensive commercially available company-level global database at present (Fuest \& Riedel 2012; Dharmapala 2014).\par
We establish the GON to clarify ownership structures of MNCs. This network is a weightless directed graph $G\in{V, E}$ composing a set of nodes $V$ and a set of links $E$ on the basis of ownership information included in the Orbis database. A node $u$ represents a company or an individual. The link $e\in{E}$ is a set of ordered pair $e = (u, v)$ (where $u, v\in{V}$) represents a shareholding relationship; $u$ is a company, and $v$ is its shareholder; self-loops are excluded ($u\neq{v}$).\par

\subsection*{Fortune Global 500}
The focus of this analysis is on MNCs listed in the Fortune Global 500. Fortune magazine complies and publishes Fortune Global 500, which are the most 500 largest companies that sell worldwide based on their total revenues (Fortune 2015). In 2017, listed MNCs have sold products worth \$30 trillion and hired 67.7 million people worldwide, so it can be said that they are representatives of huge MNCs.\par
Previous studies on intermediate companies have been limited to MNCs in the US and Germany (Collins and Shackelford 1998; Mintz and Weichenrieder 2010). In our study, we are not focusing on the locations of MNCs but on their sizes because large MNCs are likely to shift their profits to other jurisdictions. \footnote {International profit shifting costs the associated costs of constructing a complicated ownership structure and receiving expert advice familiar with the tax system of each jurisdiction.}. Therefore, we target 480 MNCs with the the largest sales in the world that have the records of their headquarters in the Orbis database and that are listed in the Fortune Global 500.\par
\begin{figure}[h!]
\centering
 \includegraphics[width=6.0cm, angle=270]{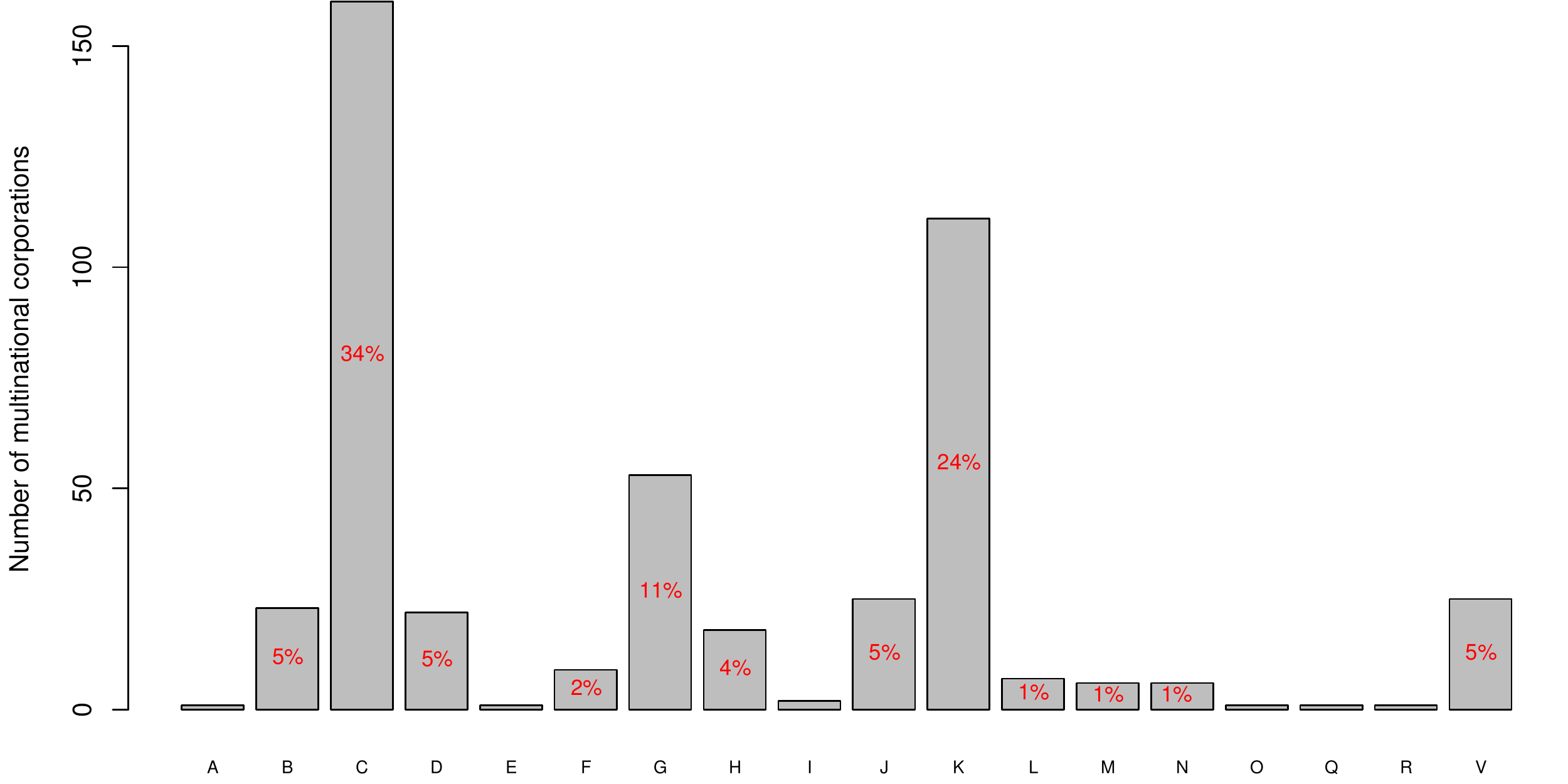}
 \caption{{\bf SIC of the headquarters of the MNCs subject to this analysis.} A is Agriculture, Forestry, and Fishing. B is Mining and Quarrying. C is Manufacturing. D is Electricity, Gas, steam, and Air conditioning supply. E is Water supply, Sewerage, Waste management, and Remediation activities. F is Construction. G is Wholesale and Retail trade, and Repair of motor vehicles and motorcycles. H is Transportation and Storage. I is Accommodation and Food service activities. J is Information and Communication. K is Financial and Insurance activities. L is Real estate activities. M is Professional, Scientific, and Technical activities. N is Administrative and Support service activities. O is Public administration and Defense, and Compulsory social security. Q is Human health and Social work activities. R is Arts, Entertainment, and Recreation. V is no information in the dataset.}
 \label{headquarter_industry}  
\end{figure}

Figure \ref{headquarter_industry} shows SIC of MNCs listed in the Fortune Global 500 (subject to our analysis) according to the statistical classification of economic activity in the European Community Second Edition (NACE Rev. 2). Information on SIC is based on the Orbis database. Many MNCs listed in the Fortune Global 500 operate in the manufacturing industry or financial insurance industry.\par

\begin{figure}[h!]
\centering
  \includegraphics[width=6.0cm, angle=270]{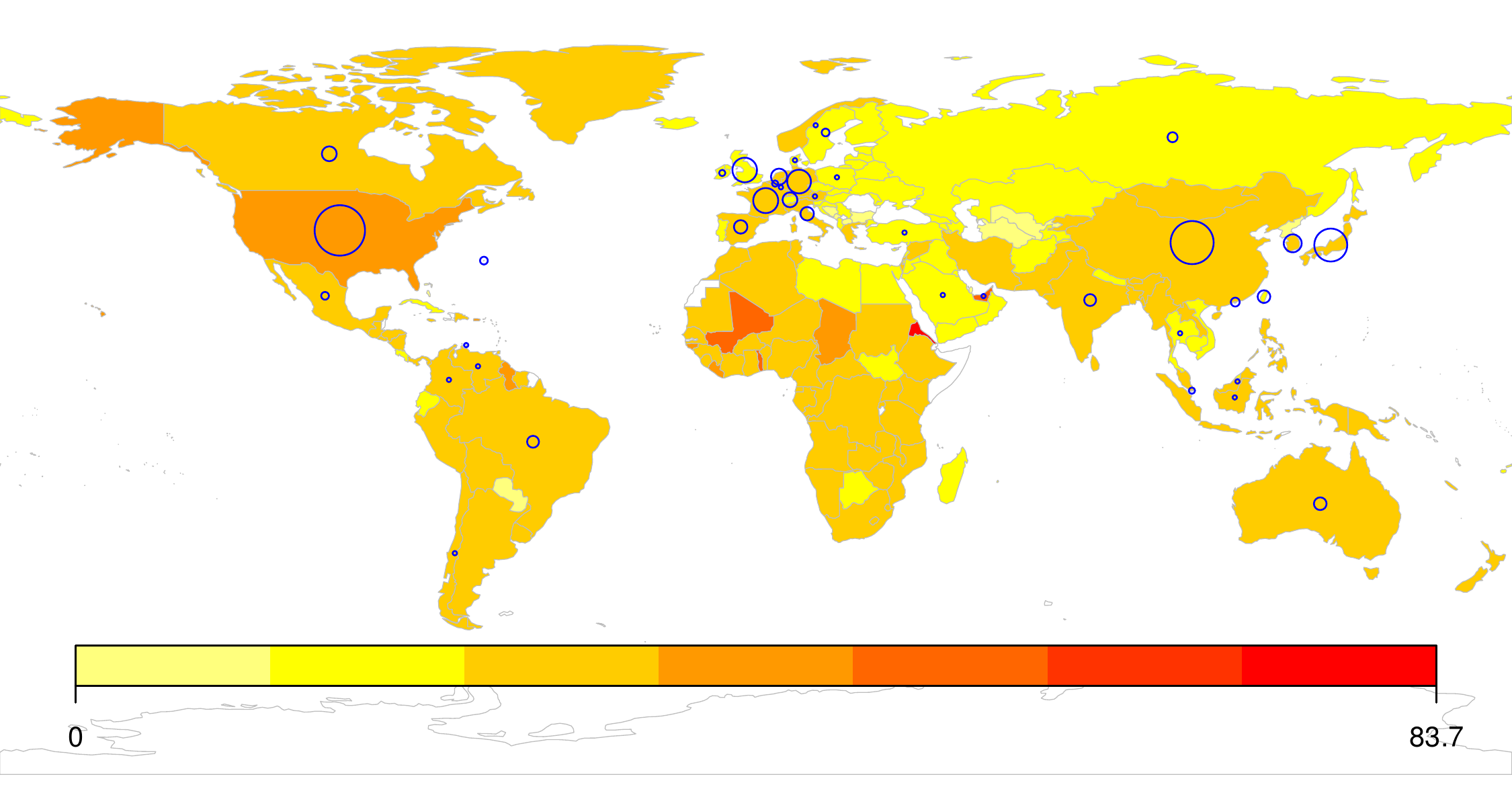}
  \caption{{\bf Geographical distribution of the headquarters of the MNCs subject to this analysis and statutory corporate tax rate.} The size of the circles represents the number of headquarters of the MNCs subject to this analysis and the depth of the color indicates the statutory corporate tax rate, by the jurisdiction.}
  \label{headquarter_tax}  
\end{figure}

Figure \ref{headquarter_tax} represents the locations of the headquarters of the MNCs based on the analysis of the corporate tax rate of each jurisdiction on the world map. Almost half of the MNCs have their headquarters in the US, China, and Japan that have relatively high statutory corporate tax rates. It is suggested that the locations of the MNCs' headquarters have nothing to do with the height of the statutory corporate tax rate.

\section*{Model}
We propose a model to identify intermediate companies, which are key companies, that are likely to play an important role in MNCs reallocation of their profits for tax purpose, namely, international profit shifting\footnote{To confirm that a company really involves in international profit shifting or international tax avoidance, we need to consider the legitimacy of the company's business (Okamura and Sakai 2018).}. Our model consists of two parts: ``holding" company and ``conduit" company centralities and hierarchical identification algorithm. We propose the ``holding" company and ``conduit" company centralities on the basis of ``sink" and  ``conduit" outward centralities, which were proposed and proved to be effective in identifying jurisdictions used for tax purpose by Garcia-Bernardo et al. (2017). Unlike the ``sink" and  ``conduit" outward centralities, the ``holding" company and ``conduit" company centralities do not need financial information and are calculated hierarchically when identifying the key companies.\par
The structure of the Model section is as follows: In the Key companies subsection, we make the concept of the key companies clear and mention that key companies have hierarchical relationships. In the Centralities for Jurisdictions subsections, we briefly introduce the ``sink" and ``conduit" outward centralities, which the ``holding" company and ``conduit" company centralities are based on. In the Calculation of the Amount of the Capital subsection, we point out that the ``sink" and ``conduit" outward centralities have the difficulty of identifying affiliates used for tax purpose in terms of financial information and explain our idea for overcoming the difficulty. In the Centralities for Affiliates subsection, we propose the ``holding" company and ``conduit" company centralities, which are able to identify the key companies, using the idea. In the Hierarchical Identification Algorithm subsection, we describe how the ``holding" company and ``conduit" company centralities are implemented for identifying the key companies. 

\subsection*{Key Companies}
In this analysis, MNC's affiliates are categorized following the classification of Garcia-Bernardo et al. (2017), which classified jurisdictions used for tax purposes into two types: ``sink" and ``conduit." The ``sink" jurisdiction attracts and parks large capital for mitigating corporate tax, whereas the ``conduit" jurisdiction allows MNCs to route their capital from other (especially, high tax) jurisdictions to the ``sink" jurisdictions without withholding tax\footnote{Withholding tax is usually imposed on the dividends when a firm pays its dividends to another firm in another jurisdiction.}. Corresponding with their classification, we categorize MNCs' affiliates used for tax purposes into two types: ``holding" and ``conduit." The ``holding"  company is defined as the affiliate that parks large capital of other affiliates, whereas the ``conduit" company is defined as the affiliate that allows MNCs to route their capital from other affiliates to the ``holding" companies \footnote{We call an affiliate that corresponds to ``sink" as ``holding" because such affiliate is usually called as a holding company by experts of international taxation (Picciotto 1992; Ginsberg 1994).}.\par
From the viewpoint of the ownership relations, we consider that key company satisfies two conditions. First, in the realm of international taxation, holding companies are classified into three, depending on their locations: home country, host country, and third country (Mintz and Weichenrieder 2010). The third country is located in a different jurisdiction from its headquarters and one of its underlying affiliates, is thought to be a key company because this type is likely to be used for international profit shifting. We call this condition as ``the condition of the third country type." Next, From the viewpoints of affiliates' position in the ownership structures, we assume that ``holding" company and ``conduit" company make a hierarchical relationships (Maine and Nguyen 2017). In other words, ``holding" company is tied with a  ``conduit" company by ownership links.\par

\subsection*{Centralities for Jurisdictions}
To identify jurisdictions functioning as ``sink" and ``conduit," Garcia-Bernardo et al. (2017) proposed the ``sink" and ``conduit" outward centralities. The ``sink" and ``conduit" outward centralities measured the amount of foreign capital flowing in an ownership network. Before explaining our new centralites, that are the ``holding" company and ``conduit" company centralities, we briefly introduce the ``sink" and ``conduit" outward centralities, which the ``holding" company and ``conduit" company centralities are based on.

\subsubsection*{Sink Centrality}
The ``sink" centrality measured the amount of foreign capital parked in a jurisdiction and identified the jurisdictions attracting a lot of capital from other jurisdictions. The ``sink" centrality $S_j$ evaluated the amount of capital parked in a jurisdiction $j$ as follows: First, they calculated the difference between the total amount of foreign capital entering into the jurisdiction $j$ and the total amount of foreign capital leaving the jurisdiction $j$. Next, the difference was divided by the total amount of capital flowing in an entire ownership network. Finally, the centrality was normalized using the economic scale of the jurisdiction $j$.

\begin{equation}
S_j = \frac{V_j^{in}-V_j^{out}}{\sum_{i}V_i^{in}}\cdot\frac{\sum_{i}GDP_i}{GDP_j}
\end{equation}

Here, $V_j^{in}$ was the total amount of foreign capital entering the jurisdiction $j$; $V_j^{out}$ was the total amount of foreign capital leaving the jurisdiction $j$; ${\sum_{i}V_i^{in}}$ was the total amount of capital flowing in the entire ownership network. $GDP_j$ is the GDP of the jurisdiction $j$; and $\sum_{i}GDP_i$ is the total of the GDP of all jurisdictions. The GDP was assumed to represent the economic scale of a jurisdiction. The ``sink" centrality $S_j$ was normalized by an economic scale of an jurisdiction $j$ as it was supposed that the difference between the total amount of foreign capital entering a jurisdiction and those leaving the jurisdiction was proportional to the economic scale of the jurisdiction $j$. When the difference (foreign capital parked in the jurisdiction $j$) was not proportional, the ``sink" centrality $S_j$ became large and it was suggested that the jurisdiction $j$ functioned as ``sink." They supposed that a jurisdiction with its ``sink" centrality exceeding 10 functioned as ``sink," comparing with the jurisdictions empirically known as "sink."\par

\subsubsection*{Conduit Outward Centrality}
The ``conduit" outward centrality measured the amount of foreign capital passed through a jurisdiction and identified the jurisdictions allowing foreign capital to be routed\footnote{Although there was ``conduit" inward centrality, we explain only ``conduit" outward centrality because the ``conduit" company centrality is based on the ``conduit" outward centrality.}. The ``conduit" outward centrality $C_j$ evaluated the amount of foreign capital routed through a jurisdiction $j$ as follows: First, they divided the amount of foreign capital routed from other jurisdictions to the ``sink" jurisdictions through the jurisdiction $j$ by the total amount of capital in an entire ownership network. Next, the centrality was normalized using the economic scale of the jurisdiction $j$.

\begin{equation}
C_j=\frac{V_j^{pass}}{\sum_{i}V_i^{pass}}\cdot\frac{\sum_{i}GDP_i}{GDP_j}
\end{equation}

Here, $V_j^{pass}$ is the total amount of foreign capital routed through the jurisdiction $j$; and $\sum_{i}V_i^{pass}$ is the total amount of foreign capital routed in an entire ownership network. $GDP_j$ is the GDP of the jurisdiction $j$ and $\sum_{i}GDP_i$ is the total of the GDP of all jurisdictions in an entire ownership network. The GDP was assumed to represent the economic scale of a jurisdiction. The ``conduit" outward centrality $C_j$ was normalized by an economic scale of an jurisdiction $j$ as it was supposed that the total amount of foreign capital passing through the jurisdiction $j$ was proportional to the economic scale of the jurisdiction $j$. When the total amount was not proportional, the ``conduit" outward centrality $C_j$ became large and it was suggested that the jurisdiction $j$ functioned as ``conduit." They supposed that a jurisdiction with its ``conduit" outward centralities exceeding 1 functioned as ``conduit," comparing with the jurisdictions empirically known as ``conduit."\par

\subsection*{Calculation of the Amount of Foreign Capital}
In the calculation of the ``sink" centrality $S_j$ and ``conduit" outward centrality $C_j$, the amount of foreign capital was assumed to be the sum of the value multiplied operating income by shareholding ratio for each company (Garcia-Bernardo et al. 2017). Thus, financial information of each company was necessary to calculate ``sink" centrality $S_j$ and ``conduit" outward centrality $C_j$. As a preliminary survey of this analysis, we examined five MNCs: Amazon, Apple, Google, Microsoft, and Starbucks. These MNC's affiliates were playing an important role in the international tax avoidance (US Senate 2013; Maine and Nguyen 2017 et al.). As a result, we found that availability of financial information about the affiliates was 15\%. In addition, no financial information was included in the database regarding affiliates that played an important role in international tax avoidance. This suggested that it was difficult to use financial information and analyze the MNCs that involved in international tax avoidance, although previous studies have mainly used financial information of each affiliate to analyze MNCs' international profit shifting (Huizinga and Laeven 2008 et al.).\par 
We aim to analyze large MNCs such as the five MNCs. Generally, large MNCs face a higher risk when implementing international profit shifting because the total amount of tax to pay are large. To overcome the difficulty of obtaining financial information of their affiliates, we measure the amount of foreign capital focusing on substantial ownership links and assume that the greater the number of the substantial ownership links, the more the capital flows. Here, a shareholding relationship of 10\% or more is considered as a substantial ownership link. It is appropriate to use 10\% or more shareholding relationships as the proxy for capturing the amount of foreign capital for international taxation because shareholders that own 10\% or more of shares are subject to the controlled foreign company legislation (CFC rules), which is so-called anti-tax haven regime (Yoshimura 2008).\par

\subsection*{Centrality for affiliates}
To identify MNC's affiliates key companies, that are ``holding" and ``conduit" companies, we  propose the ``holding" company and ``conduit" company centralities. Since the ``holding" company and ``conduit" company centralities capture the amount of foreign capital by the number of substantial ownership links, the centralities do not need financial information such as operating income. 

\subsubsection*{Holding Company Centrality}
On the basis of the ``sink" centrality $S_j$, we define ``holding" company centrality to identify MNCs' affiliates functioning as ``holding." The ``holding" company centrality measures the amount of foreign capital held by an affiliate and identify the affiliates holding a lot of foreign capital from other affiliates. The ``holding" company centrality $H_a$ evaluates the amount of foreign capital held by an affiliate $a$ as follows: First, we calculate the difference between the total number of substantial ownership links where foreign capital enters the affiliate $a$ and the total number of substantial ownership links where foreign capital leaves the affiliate $a$. Next, we divide the difference by the total number of substantial ownership links where capital flows in the GON. Finally, the centrality is normalized using the economic scale of the affiliate $a$.

\begin{equation}
H_a = \frac{k^{in}_a-k^{out}_a}{\sum_{b\in{M}}k^{in}_b} \cdot \frac{\sum_{b\in{M}}(k^{in}_b+k^{out}_b)}{k^{in}_a+k^{out}_a}
\end{equation}

Here, $M$ is the MNC that controls affiliates; $a$ is an affiliate of the MNC $M$, $k^{in}_a$ is the number of the in-degree of the affiliate $a$ (the total number of substantial ownership links where foreign capital enters the affiliate $a$); $k^{out}_a$ is the number of the out-degree of the affiliate $a$ (the total number of substantial ownership links where foreign capital leaves the affiliate $a$). $k^{in}_a + k^{out}_a$ is the number of the degree of the affiliate $a$ (the total number of ownership links tied with the affiliate $a$) and is assumed to represent the economic scale of the affiliate $a$. The ``holding" company centrality $H_a$ is normalized by an economic scale of an affiliate $a$ as it is supposed that the difference between the amount of foreign capital entering the affiliate $a$ and those leaving the affiliate $a$ is proportional to the economic scale of the affiliate $a$. When the difference is not proportional, the ``holding" centrality $H_a$ becomes large and it is suggested that the affiliate $a$ is likely to function as ``holding." In this analysis, we suppose that an affiliate with its ``holding" company centrality exceeding 0 has the possibility of identification as a "holding" company.

\subsubsection*{Conduit Company Centrality}
On the basis of ``conduit" centrality $C_j$, we define ``conduit" company centrality to identify MNCs' affiliates functioning as ``conduit." The ``conduit" company centrality measures the amount of foreign capital passed through an affiliate and identifies the affiliates allowing foreign capital to be routed. The ``conduit" company centrality $T_a$ evaluates the amount of foreign capital routed to the ``holding" companies through an affiliate $a$ as follows: First, we divide the total number of substantial ownership links where foreign capital passes through the affiliate $a$ and flows to ``holding" companies by the total number of substantial ownership links where capital flows in the GON. Then, the centrality is normalized using the economic scale of the affiliate $a$.

\begin{equation}
T_a = \frac{k^{in}_a}{\sum_{b\in{M}}k^{in}_b \cdot k^{out}_b} \cdot \frac{\sum_{b\in{M}}(k^{in}_b+k^{out}_b)}{k^{in}_a+k^{out}_a}
\end{equation}

Here, $M$ is the MNC that controls affiliates; $a$ is an affiliate of the MNC $M$; $k^{in}_a$ is the number of the in-degree of the affiliate $a$ (the total number of substantial ownership links where foreign capital passes through the affiliate $a$ and flows to ``holding" companies); $k^{in}_a+k^{out}_a$ is the number of the degree of the affiliate $a$ (the total number of substantial ownership links tied with the affiliate $a$) and is assumed to represent the economic scale of the affiliate $a$. The ``conduit" company centrality $T_a$ is normalized by an economic scale of an affiliate $a$ because it is supposed that the total number amount of foreign capital passing through the affiliate $a$ is proportional to the economic scale of the affiliate $a$. When the total amount was not proportional, the ``conduit" centrality $T_a$ becomes large and it is suggested that the affiliate $a$ is likely to function as a ``conduit." In this analysis, we suppose that an affiliate with its ``conduit" company centrality exceeding 0 has the possibility of identification as a "conduit" company.

\subsection*{Hierarchical Identification Algorithm}

\begin{figure}[h!]
\centering
  \includegraphics[width=7.5cm]{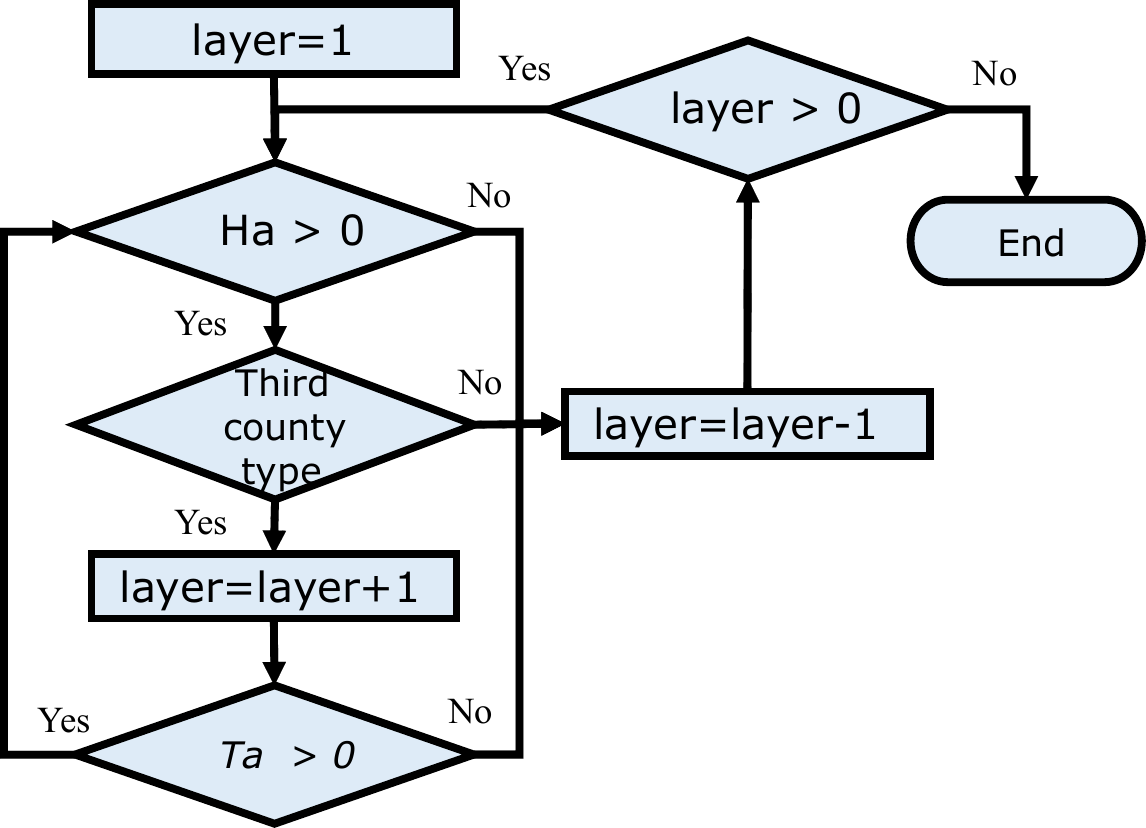}
\caption{{\bf Flow chart of our model.} $a$ indicates a layer. First of all, the ``holding" company centrality $H_a$ is calculated for affiliates in the first ownership layer. Of it exceeds 0, the ``conduit" company centrality $T_a$ is calculated for affiliates held by the affiliate with the ``holding" company centrality $H_a$ exceeding 0.}
\label{Algorithms}
\end{figure}

Assuming that a ``holding" company and a ``conduit" company make hierarchical relationships (see the Key Companies subsection), we calculate the ``holding" company centrality $H_a$ and ``conduit" company centrality $T_a$ following the algorithms indicated in Figure \ref{Algorithms}. We start the calculation from the first ownership layer. First, we calculate the  ``holding" company centrality $H_a$ of an affiliate (e.g. {\it abc affiliate}) in the first ownership layer. If the ``holding" company centrality $H_a$ exceeds 0 and , then we go to the second ownership layer and calculate the ``conduit" company ownership centrality of affiliates held by {\it abc affiliate}. If there is a affiliate (e.g. {\it def affiliate}) whose ``conduit" company centrality $T_a$ exceeds 0, {\it abc affiliate} is identified as a ``holding" company and {\it def affiliate} is identified as a ``conduit" company. Next, we calculate the ``holding" company centrality $H_a$ of {\it def affiliate}. If the ``holding" company centrality exceeds 0 and {\it def affiliate} satisfies the condition of the third country type, {\it def affiliate} is identified as a ``holding and " conduit company because {\it def affiliate} is also a ``conduit" company toward {\it abc affiliate}. Following this algorithm, the ``holding" company and ``conduit" company centralities are calculated for all affiliate of each MNC.\par

\begin{figure}[h!]
\centering
  \includegraphics[width=7.5cm]{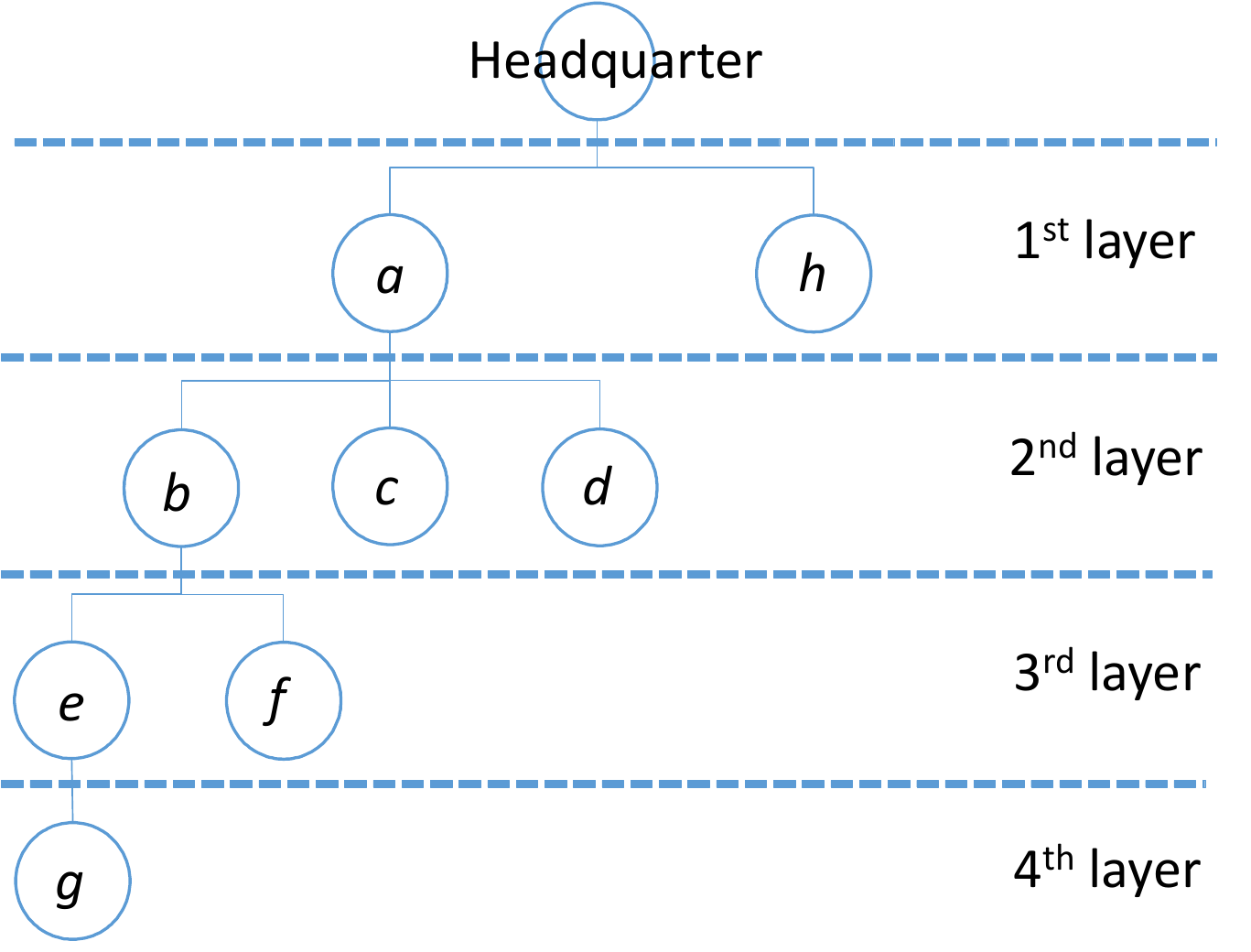}
\caption{{\bf Ownership structure of MNC $M$.} the circles indicate the headquarter and its affiliates; the lines indicate the substantial ownership links; the labels indicate the name of the affiliates; and the dotted lines indicate the division the ownership layer.}
\label{example}
\end{figure}

Figure \ref{example} shows a ownership structure of a MNC $M$ as an example. We use the MNC $M$ for the explanation of the algorithm. The algorithm is applied to the MNE $M$ as follows: First, we calculate the ``holding" company centrality $H_a$ of the affiliate $a$ in the ownership first layer. Since the ``holding" company centrality $H_a$ of the affiliate $a$ exceeds 0 and the affiliate $a$ satisfies the conditions of the third country type, the ``conduit" company centrality is calculated regarding each affiliate tied with the affiliate $a$ by the substantial ownership links, namely, the affiliates $b$, $c$, and $d$. Within these three affiliates $b$, $c$, and $d$, the ``conduit" company centrality $T_b$ of the affiliate $b$ exceeds 0. Thus, the affiliate $a$ is identified as a ``holding" company toward the affiliates $b$, $c$, and $d$, while the affiliate $b$ is identified as a ``conduit" company toward the affiliate $a$. Second, we calculate the ``holding" company centrality $H_b$ of the affiliate $b$ because its ``conduit" company centrality $T_b$ exceeds 0. Since the ``holding" company centrality $H_b$ of the affiliate $b$ exceeds 0 and the affiliate $b$ satisfies the conditions of the third country type, the ``conduit" company centrality is calculated regarding each affiliate tied with the affiliate $b$ by the substantial ownership links, namely, the affiliates $e$ and $f$. Within these two affiliates $e$ and $f$, the ``conduit" company centrality $T_e$ of the affiliate $e$ exceeds 0. Thus, the affiliate $b$ is identified as a ``holding" company toward the affiliates $e$ and $f$, while the affiliate $e$ is identified as a ``conduit" company toward the affiliate $b$. Because the affiliate $b$ is not only a ``holding" company toward the affiliates $e$ and $f$ but also a ``conduit" company toward the affiliate $a$, we regard the affiliate $b$ as a ``holding and conduit" company. Third, we calculate the ``holding" company centrality $H_e$ of the affiliate $e$ because its ``conduit" company centrality $T_e$ exceeds 0. Although the affiliate $e$ satisfies the condition of the third country type, the affiliate $e$ is not identified as a ``holding" company because its ``holding" centrality $H_e$ does not exceed 0. There is no other affiliates whose ``conduit" company centrality exceeds 0 within the affiliates tied with the affiliates $b$ by the substantial ownership links, namely, the affiliates $e$ and $f$. Fourth, we return to the first ownership layer and calculate the ``holding" centrality of the affiliate $h$. The ``holding" centrality $H_h$ of the affiliate $h$ does not exceeds 0 and there is no other affiliates in the first ownership layer. Finally, we finish our identification. We calculate the ``holding" company and ``conduit" company centralities following this algorithm and identify the key companies of the MNCs listed in the Fortune 500 (subject to this analysis).\par

\section*{Results and Discussion}
We demonstrate the characteristics, structure, and communities of the GON. Regarding five MNCs whose affiliates play an important role in the international tax avoidance, we confirm the validity of our model by verifying how our model identifies the affiliates in percentage. Of the 361,166 affiliates holding (directly or indirectly) by MNCs listed in the Fortune Global 500 (subject to this analysis), our model identifies 6223 affiliates as the key companies that are at risk of the international profit shifting. With the behavior of the key companies, it is revealed how key companies distribute throughout jurisdictions and in the GON. We also point out that the results have certain bias due to the bias of the recording ratio of the database.

\subsection*{Verification of Proposed Model}
We surveyed five MNCs: Amazon, Apple, Google, Microsoft, and Starbucks. We checked the extent our model identifies affiliates that play an important role in international tax avoidance. The reason selecting these five MNCs is that we can identify the affiliates they use for the international tax avoidance (UK Parliament 2013; US Senate 2013; Fuest et al. 2013; Gravelle 2013; Kleinbard 2013; Connell 2014; Maine and Nguyen 2017 et al.).\par

\begin{figure}[ht]
\centering
\subfigure[Google affiliates]{
\includegraphics[width=5.2cm]{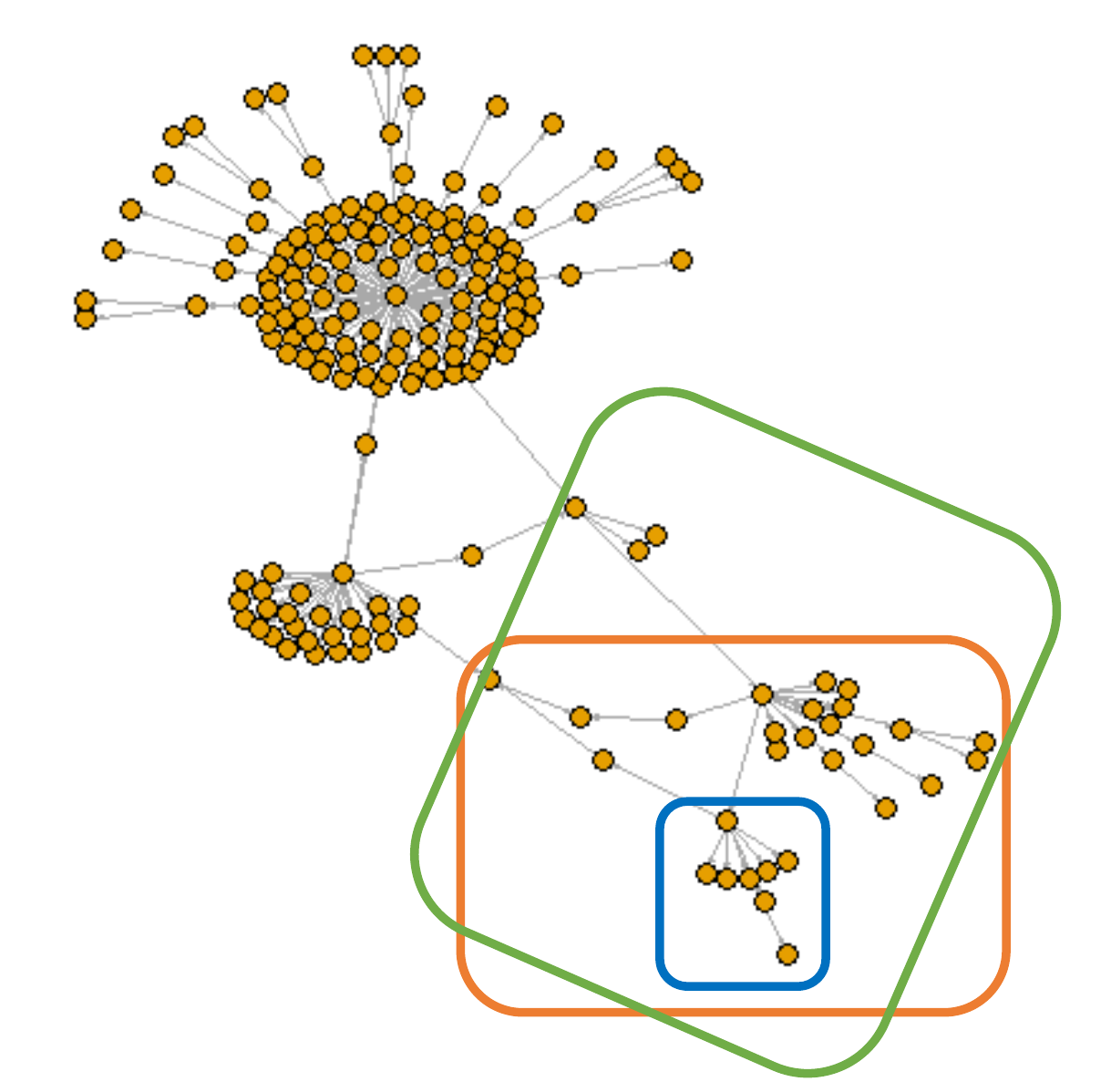}
}
\subfigure[First layer]{
\includegraphics[width=5.2cm]{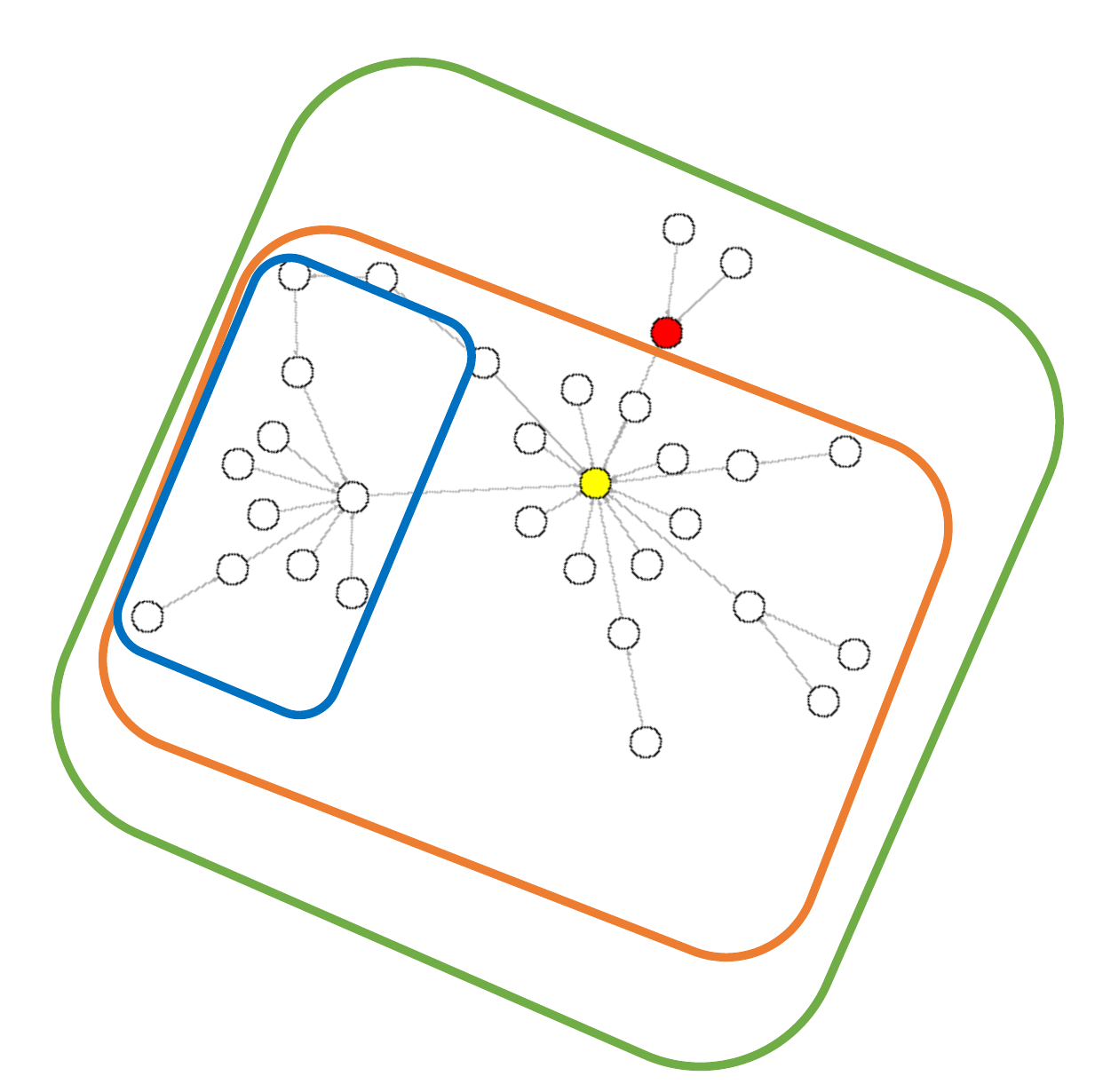}
}\par
\subfigure[Second layer]{
\includegraphics[width=4.8cm]{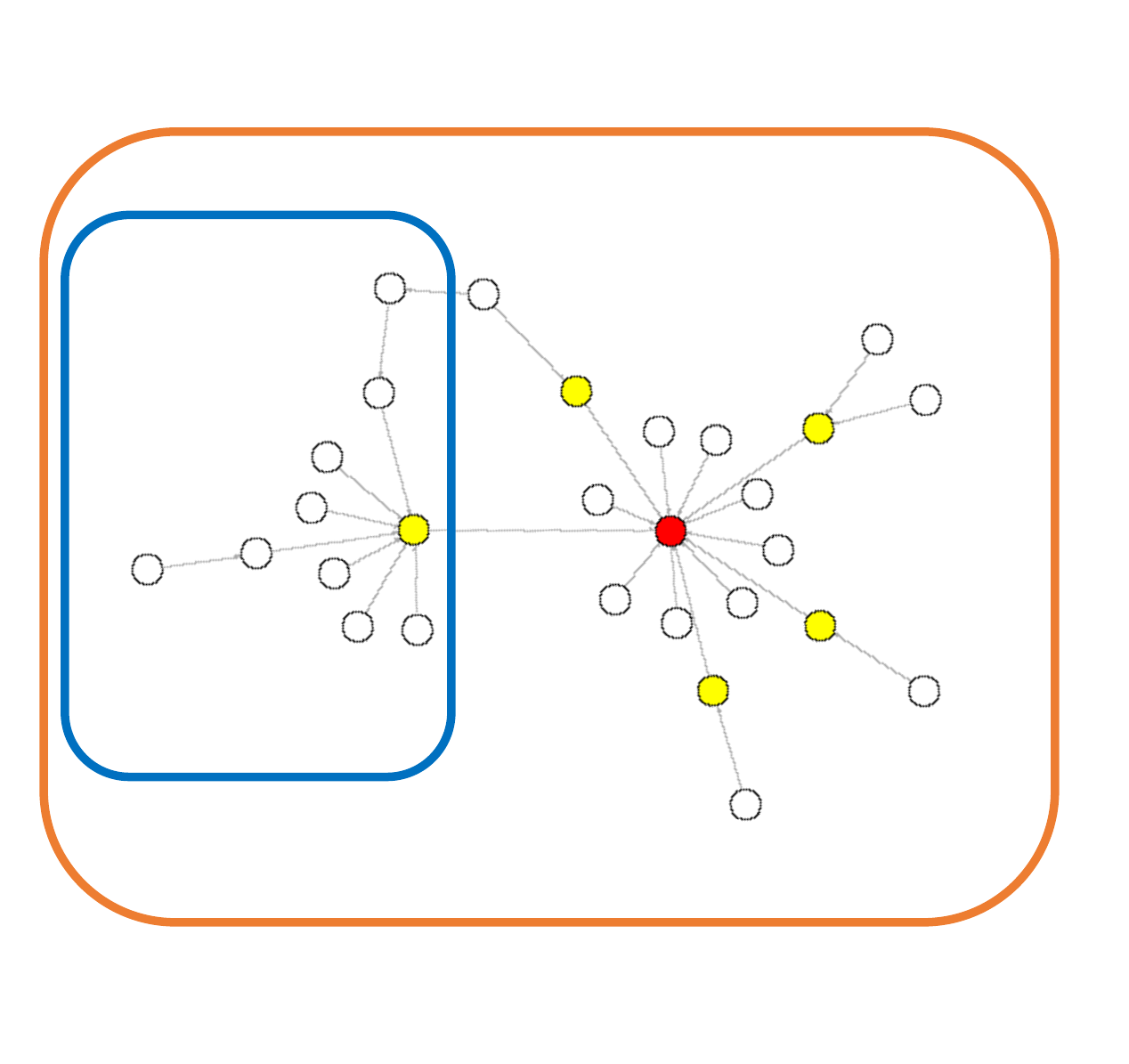}
}
\subfigure[Third Layer]{
\includegraphics[width=4.2cm]{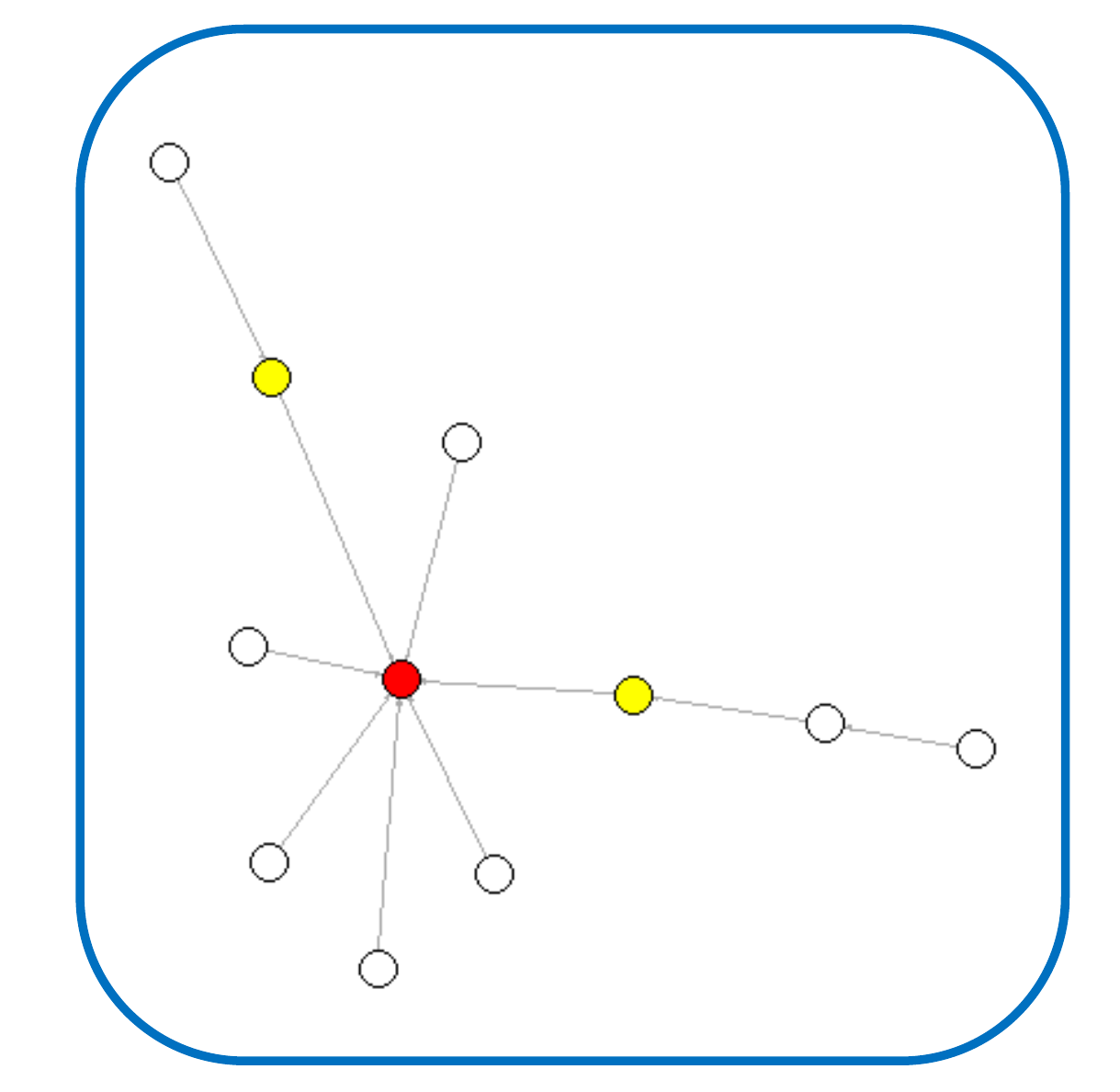}
}
\caption{{\bf Example of Google.} (a) shows all Google's affiliates. (b)-(d) show the result of calculation in the first-third layers, respectively. A company whose holding centrality $H$ exceeds 0 is painted in red and a company whose conduit centrality $C$ exceeds 0 is painted in yellow. Green, orange, and blue rectangles show the first-third layer, respectively.}
\label{fig:google}       
\end{figure}

Using Google, which actually exists, as an example, we describe how the algorithms identify key companies, that are ``holding" and ``conduit" companies, by showing an important part of the result. Figure \ref{fig:google} (a) shows 190 affiliates that are directly or indirectly tied with Google LLC, which is the headquarter of Google, by the substantial ownership links\footnote{We considered Google LLC as the headquarter of Google because the Orbis database 2015 does not include the information of Alphabet Inc, which was established in 2015.}.  We calculated ``holding" company and ``conduit" company centralities for the 190 affiliates, following the hierarchical identification algorithm. In Figure \ref{fig:google} (b), we show that the first ownership layer is indicated by a green rectangle; one affiliate is painted in red because its ``holding" company centrality exceeds 0, and one affiliate is painted in yellow because its ``conduit" company centrality exceeds 0. Therefore, one affiliate painted in red is identified as a ``holding" company. Figure \ref{fig:google} (c) shows the result of the calculation in the second ownership layer; this is indicated by an orange rectangle. One affiliate is painted in red because its ``holding" company centrality exceeds 0, and  five affiliates are painted in yellow because their ``conduit" company centralities exceed 0. The one affiliate painted in red in the second ownership layer is identified as a ``holding and conduit"company because its ``conduit" company centrality exceeds 0 in the first ownership layer. On the other hand,  in the second ownership layer, only one affiliate is identified as a ``conduit" company because four affiliates do not satisfy the conditions of the third country type, even though their ``conduit" company centralities exceed 0. Figure \ref{fig:google} (d) shows the result of the calculation in the third ownership layer indicated by a blue rectangle. One affiliate is painted in red because its ``holding" company centrality exceeds 0, and two companies are painted in yellow because their ``conduit" company centralities exceed 0. The one affiliate painted in red in the third ownership layer is identified as a ``holding and conduit" company because its ``conduit" company centrality exceeds 0 in the second ownership layer. On the other hand, no affiliate is identified as a ``conduit" company in the third ownership layer because no company satisfies the conditions of the third country type, even though their ``conduit" company centralities exceed 0. The hierarchical identification algorithm works in this way regarding Google.

Table \ref{tab:result} summarizes the percentage of the five MNCs for the application of our model and their affiliates that play an important role in international tax avoidance. The column of  ``Affiliates" shows the number of affiliates; the columns of  ``Holding," ``Holding \& Conduit," and ``Conduit" show the number of affiliates identified as the key companies; and the column of  ``Reported" shows the ratio these key companies and affiliates play in the international tax avoidance. As the column of “Reported” showed, although the ratio varied from 20\% to 100\%, depending on MNCs, we confirmed the validity of our model because the identified key companies included affiliates that played the most important role in the international profit shifting of these five MNCs.

\begin{table}[h]
 \centering
  \caption{Number of key companies identified.}
   \begin{tabular}{cccccc}
\toprule
    MNCs & Affiliates & Holding & Holding \& Conduit & Conduit & Reported \\
\cmidrule(lr){1-6}
    Amazon & 312 & 2 & 0 & 2 & 25\%\\
    Apple & 152 & 1 & 1 & 0 & 50\%\\
    Google & 190 & 3 & 2 & 0 & 20\%\\
    Microsoft & 266 & 8 & 1 & 1 & 20\%\\
    Starbucks & 274 & 1 & 1 & 0 & 100\%\\ 
\bottomrule
  \end{tabular}
  \label{tab:result}
\end{table}

\subsection*{Basic Characteristics of the GON}
  \begin{figure}[h!]
\centering
  \includegraphics[width=12.0cm]{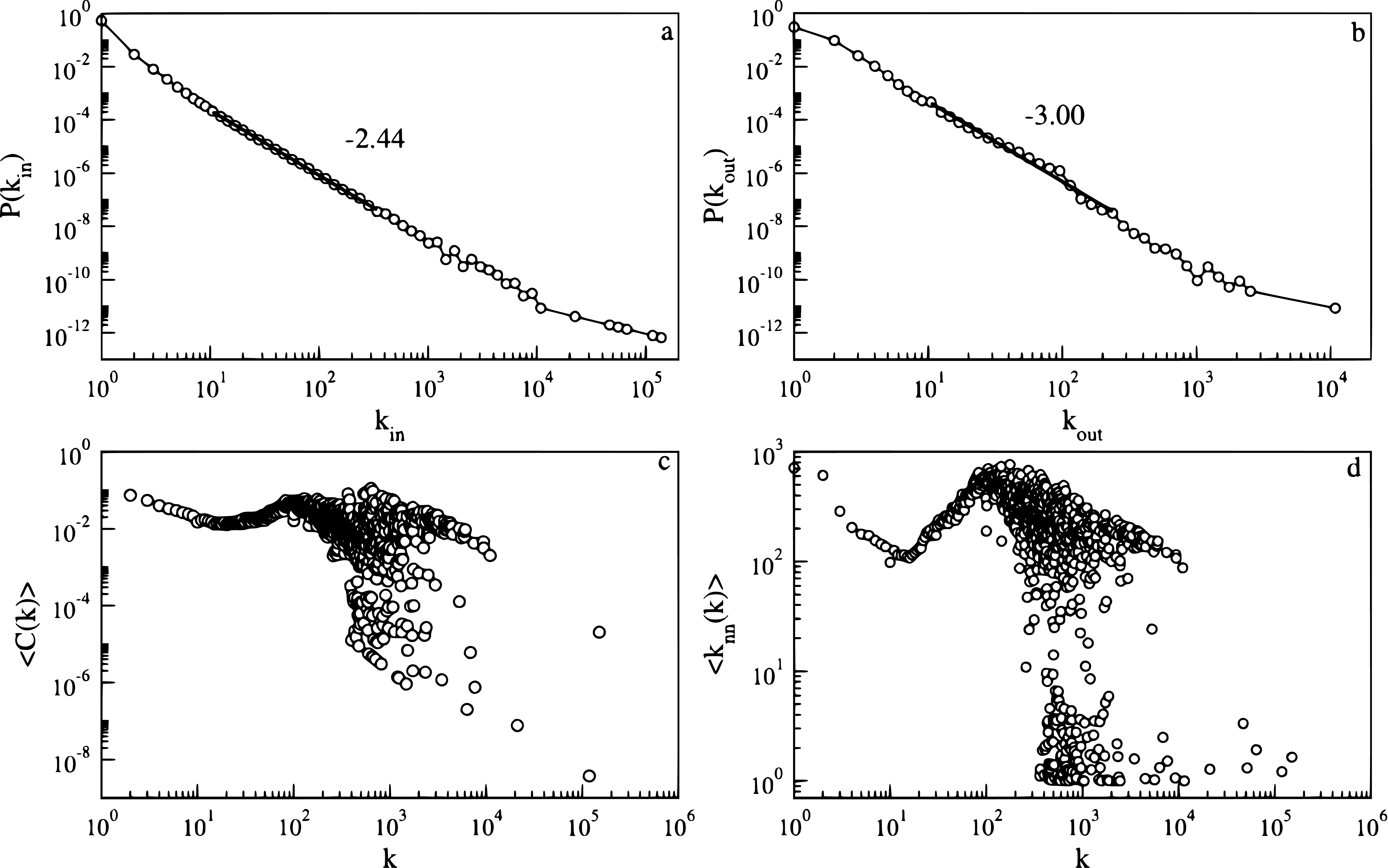}
  \caption{{\bf Structural properties of the GON.}
  \label{degree}
      {The probability density distributions $P$ for nodal (a) in-degree $k_{in}$ and (b) out-degree$k_{out}$. Solid lines indicate best power law fit to the data.  
      (c) The average clustering coefficient $\langle C(k) \rangle$ is plotted with total degree $k$. (d) The variation of average nearest neighbor degree $\langle k_{nn}(k) \rangle$ is shown as a function of nodal degree $k$.
      Logarithmic binning of the data is used in (a) and (b).}
      }
  \end{figure}

The total number of node $u$ included in the GON is 59,581,452, and the total number of link $e$ is 42,801,276. We show the structural properties of the GON in Figure~\ref{degree}. Figure~\ref{degree}~(a,b) represents the probability density distributions of nodal in- and out- degrees, respectively. On a double logarithmic scale, both the distributions have a significant straight portion of the intermediate values of the degrees. This indicates a power law decay of the degree distribution with $P(k_{in/out}) \sim k_{in/out}^{-\gamma_{in/out}}$ characterizing a scale free nature of the network. The slope of the distributions gives the values of the exponents $\gamma_{in} = 2.44 $, and $\gamma_{out} = 3.00$. The average nodal in- or out-degree  is found to be $\langle k_{in}\rangle = \langle k_{out}\rangle = 0.718$. Approximately $0.119\%$ of all links indicate cross-share relationships. The uniqueness among the neighbors of a node is measured in terms of clustering coefficients, which also measures the three-point correlation. In Figure~\ref{degree}~(c), we show the average clustering coefficient $\langle C(k) \rangle$, which decreases as the total degree $k$ increases, observed in many other real-world networks. The nodal degree-degree correlation is measured by the average nearest neighbor degree $\langle k_{nn} (k)\rangle$. If $\langle k_{nn}(k) \rangle$ decreases with $k$, the network is called disassortative; if it increases with $k$, the network is known as assortative. As can be seen from Figure~\ref{degree}~(d), $\langle k_{nn}(k) \rangle$ has an unusual shape for low $k < 200$. For low $k$, it gradually decreases until $k < 20$; then, it increases until $k < 200$. However, it decreases with very high $k$ values, indicating a disassortative nature.

 \begin{figure}[h!]
\centering
  \includegraphics[width=8.0cm]{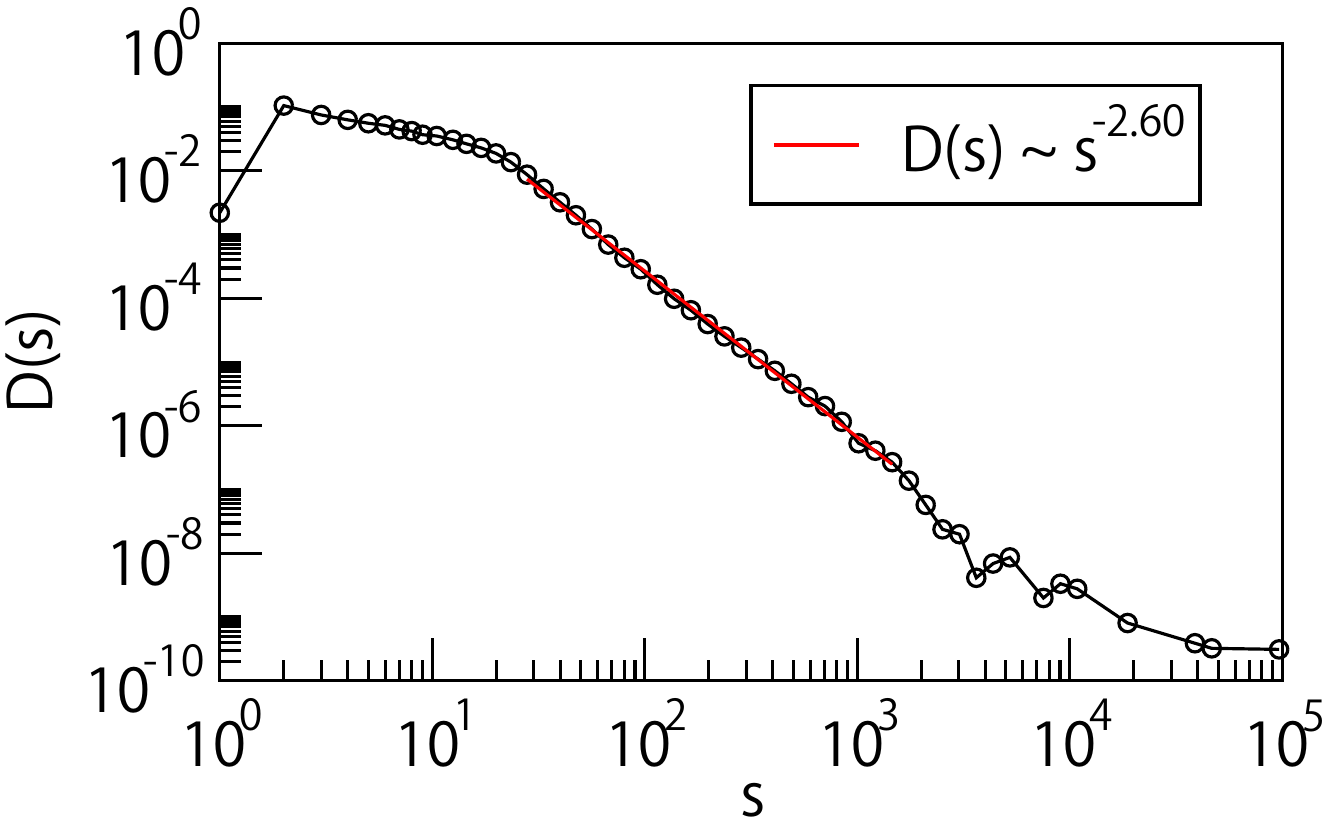}
  \caption{{\bf The probability density distributions $D(s)$ of community sizes $s$ for the GON.}
  \label{comsize}
      {The red line indicates the best power law fit to the data}
      }
  \end{figure}

Community detection is conducted on the GON using the Infomap method (Rosvall and Bergstrom 2008). This detects 363,991 communities with wide variation in their sizes $s$, where the size is measured by the total number of nodes within the community. Figure~\ref{comsize} is the probability density distribution $D(s)$ of the sizes of the communities. On the double logarithmic scale, the distribution has a considerably long straight portion reflecting a power law behavior of the form $D(s)\sim s^{-2.60}$.\par

\begin{figure}[h!]
\centering
  \includegraphics[width=8.0cm]{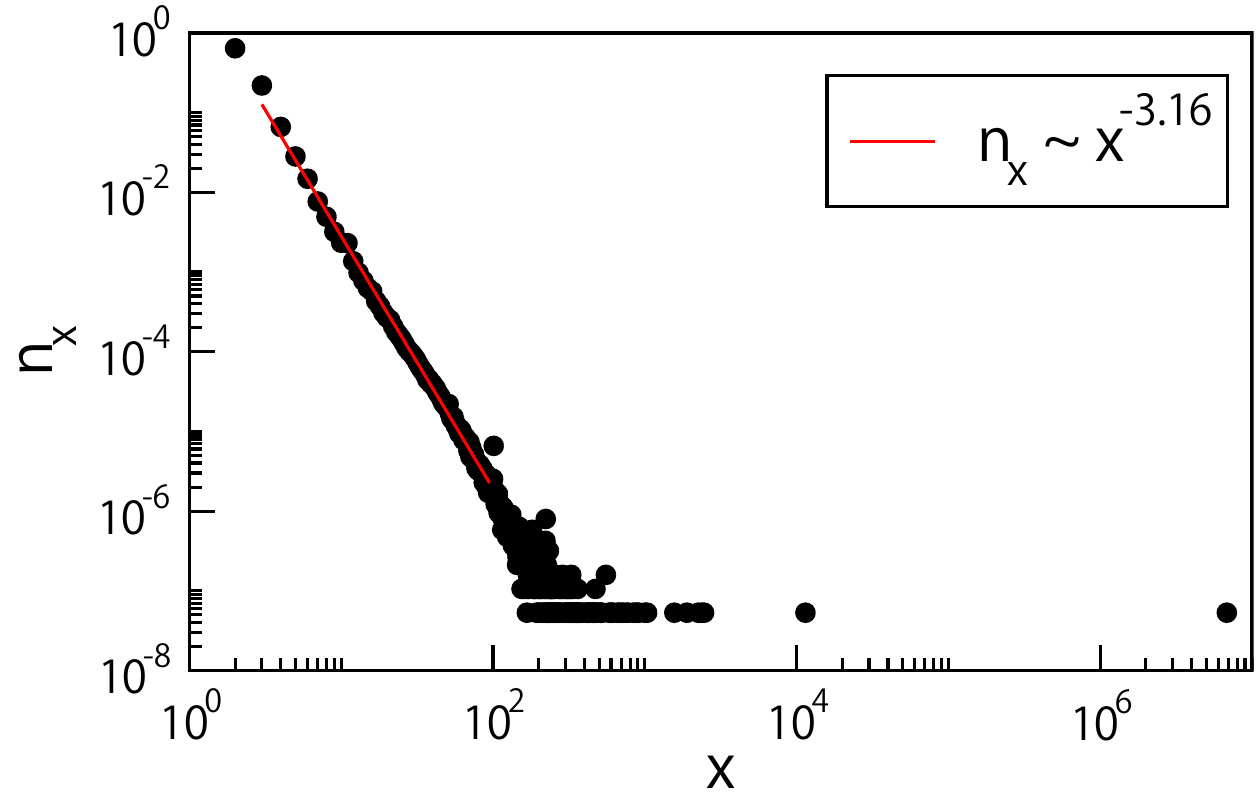}
  \caption{{\bf Distribution $n_x$ of the component sizes $x$ in the network.}
  \label{fig:comp}
      {The red line indicates the best power law fit to the data}
      }
  \end{figure}

We study the connected components when the network is viewed as an undirected network. The largest connected component of the network is known as the giant weakly connected component (GWCC). As can be seen from Figure~\ref{fig:comp}, the network consists of a very large GWCC with $N = 6,827,299$ nodes and $L = 8,367,999$ links, which is $11.459\%$ of the GON. The small components with size $x < 100$ can be fitted with a power law decay of form $n_x \sim x^{-3.16}$. In the later part of our study, we investigate the GWCC of the network. 
  
\subsection*{Bow-tie Srtucture of the GWCC}

\begin{table}[hb]
\centering
\caption{\bf Bow-tie structure: Sizes of different components.}
\begin{tabular}{c r r}
  \toprule
  Component & Companies & Ratio (\%) \\
\cmidrule(lr){1-3}
  GSCC & 2239 & 0.033  \\
  IN & 1,161,655 & 17.015  \\
  OUT & 15,514 & 0.227  \\
  TE & 5,647,891 & 82.725  \\
\cmidrule(lr){1-3}
  Total & 6,827,299 & 100 \\
  \bottomrule \noalign{\smallskip} 
\end{tabular}
\begin{flushleft}
``Ratio" refers to the ratio of the number of companies to the total number of companies in GWCC.
\end{flushleft}
\label{tab:bowtie}
\end{table}

The bow-tie structure is uncovered from the GWCC. The definitions of the different regions of the bow-tie structure are given as follows:
 \begin{itemize}
 \item The giant strongly connected component (GSCC): The largest region where any two nodes are reachable through directed paths. 
 \item IN component: The nodes from which GSCC is reachable through directed paths. 
 \item OUT component: The nodes that are reachable from the GSCC through directed paths.
 \item Tendrils (TE): The rest of the nodes in the GWCC. 
 \end{itemize}

The number of companies in each components is shown in Table~\ref{tab:bowtie}. The shape is highly asymmetric; the nodes in the strongly connected component and the OUT component are only $0.033\%$ and $0.227\%$, respectively, whereas those of the IN component is $17.015\%$. The results are consistent with the previous research analyzing the ownership network of transnational corporations in terms of the small OUT component; however, the proportion of the TE is 82.725\%, very large as compared with that of the previous research (Vitali et al. 2011). This is probably because our data are much bigger than those of the previous research.

  \begin{table}[h!]
\centering
  \caption{The distribution of the shortest distance from nodes in the IN component to those in the GSCC and from nodes in the GSCC to those in the OUT component.}
      \begin{tabular}{c rc rc}
\toprule
        & \multicolumn{2}{c}{IN to GSCC} & \multicolumn{2}{c}{OUT to GSCC}\\
\cmidrule(lr){2-3} \cmidrule(lr){4-5}
        Distance & Companies & Ratio & Companies & Ratio \\
\cmidrule(lr){1-5}
        1 & 98,342 & 8\% & 7405 & 48\% \\
        2 & 564,328 & 49\% & 5131 & 33\% \\
        3 & 296,760 & 26\% & 1897 & 12\% \\
        4 & 123,753 & 11\% & 712 & 5\% \\
        5 & 46,106 & 4\% & 269 & 2\% \\
        6 & 18,258 & 2\% & 54 & 0\% \\
        7 & 7883 & 1\% & 37 &  0\% \\
        8 & 3226 & 0\% & 4 &  0\% \\
        9 & 1563 & 0\% & 2 &  0\% \\
        10 & 705 & 0\% & 3 &  0\% \\
        11 & 319 & 0\% & - &  0\% \\
        12 & 183 & 0\% & - &  0\% \\
        13 & 65 & 0\% & - &  0\% \\
        14 & 37 & 0\% & - &  0\% \\
        15 & 33 & 0\% & - &  0\% \\
        16 & 34 & 0\% & - &  0\% \\
        17 & 13 & 0\% & - &  0\% \\
        18 & 16 & 0\% & - &  0\% \\
        19 & 23 & 0\% & - &  0\% \\
        20 & 8 & 0\% & - &  0\% \\
\cmidrule(lr){1-5}
        Total & 1,161,655 & 100\% & 15, 514 & 100\% \\
\bottomrule
      \end{tabular}
  \label{tab:route}
\end{table}

Table \ref{tab:route} shows the distribution of the shortest distance from the node on the IN component to the node on the GSCC and the shortest distance from the node on the GSCC to the node on the OUT component. Nearly 80\% of the nodes belonging to the IN component had a distance of 2-3 to the GSCC, and 80\% of the nodes that belong to the OUT component had a distance of 1-2 from the GSCC. The distances from the IN component to the GSCC are longer than those from the GSCC to the OUT component. The shape of the bow-tie structure is different from those shown in other networks (Vazquez et al. 2002; Csete M and Doyle J 2004; Chakraborty et al. 2018 et al.).

\subsection*{Identification of Key Companies}

  \begin{figure}[h!]
\centering
  \includegraphics[width=6.0cm, angle=270]{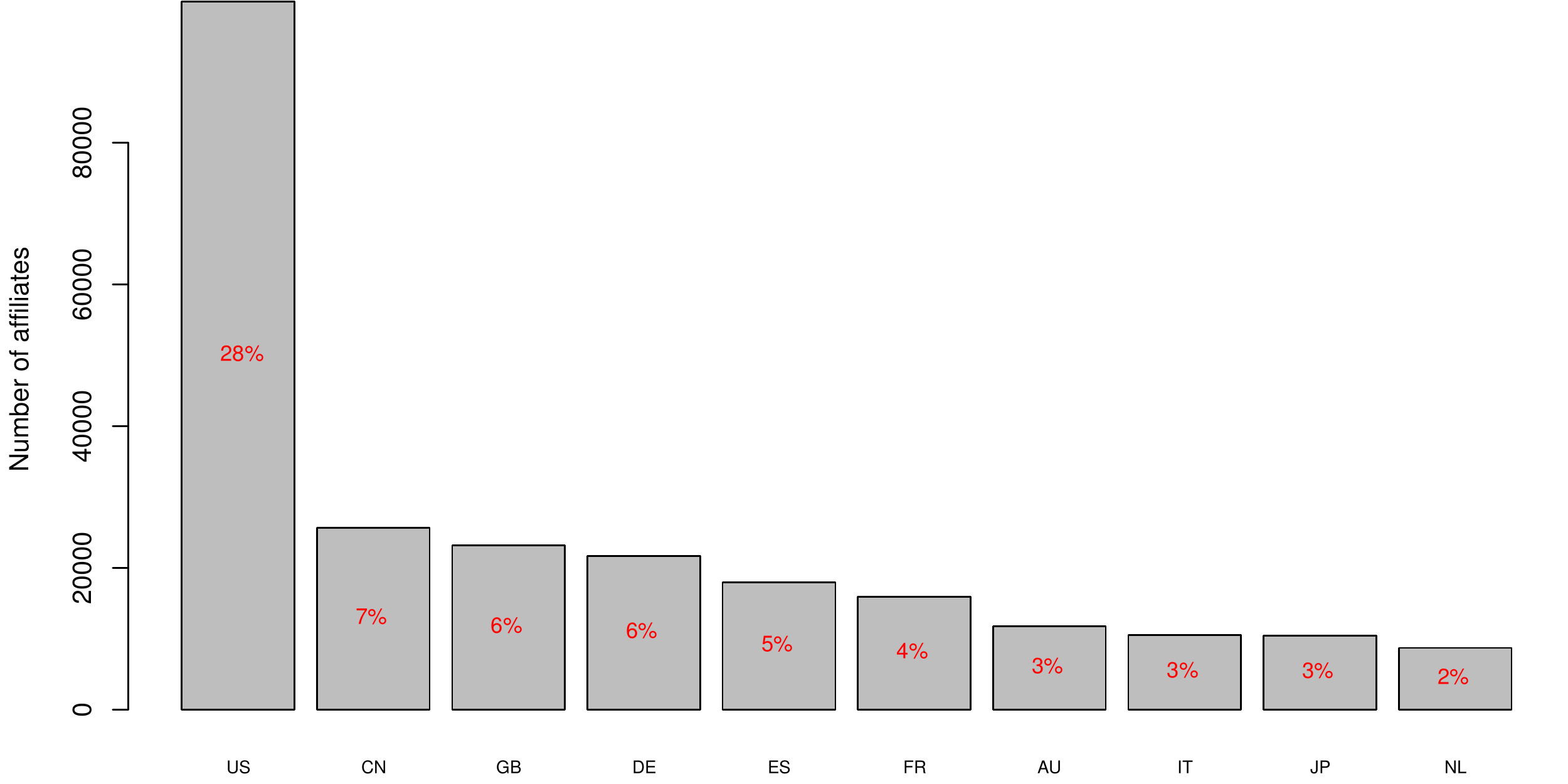}
  \caption{{\bf Top 10 Jurisdictions where affiliates are located.}
  \label{affiliates_jurisdiction}
The top 10 jurisdictions of the affiliates of the MNCs subject to this analysis. The horizontal axis indicates the locations, the vertical axis indicates the number of affiliates, and the labels indicate the percentage accounting for the total number of the affiliates. CN is mainland China; GB is the UK; DE is Germany; ES is Spain; FR is France; AU is Austria; IT is Italy; JP is Japan; and NL is the Netherlands.}
     \end{figure}

A total of 361,166 companies were extracted from the GON as affiliates of the MNCs listed in the Fortune Global 500 (subject to this analysis). Figure \ref{affiliates_jurisdiction} shows the geographic distribution of the affiliates. About 30\% of the affiliates are located in the US, and most of the other affiliates are located in Europe and East Asia. This might be because many headquarters of the MNCs listed in the Fortune 500 (subject to this analysis) are located in those jurisdictions and those jurisdictions have large markets. We also point out that the bias included the dataset might cause the locations of the affiliates to concentrate on certain areas, that are the US, Europe, and East Asia.\par

  \begin{figure}[h]
\centering
  \includegraphics[width=10.0cm]{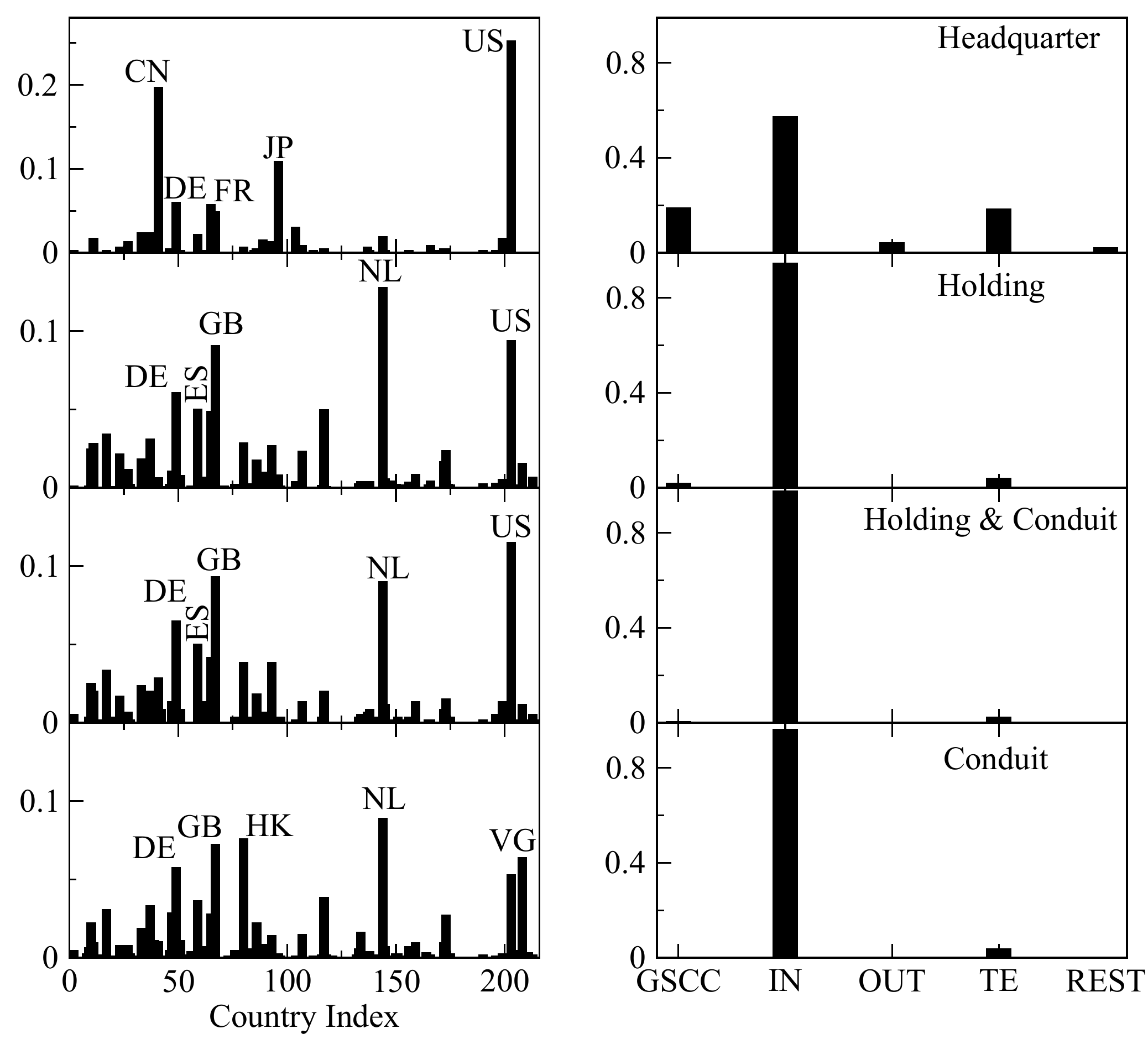}
  \caption{{\bf Distribution of the headquarters and the key companies of the MNCs subject to this analysis.}
    The left bar plots show the geographical distributions, while the right bar plots show the distributions in the bow-tie structure of the GWCC of the GON. The bar plots of both sides show regarding the headquarters, the ``holding" companies, the ``holding and conduit" companies, and conduit" companies from top to bottom. REST stands for the component except the GWCC.}
  \label{distribution}
  \end{figure}

Our model identified 3316 affiliates as ``holding", 1596 affiliates as ``holding and conduit" and 1311 affiliates as ``conduit."  The left bar plots of Figure \ref{distribution} show the geographical distributions of the headquarters and the key companies of the MNCs listed in the Fortune Global 500 (subject to this analysis). First bar plot shows regarding the headquarters. The US, China (CN), and Japan (JP) are the top three jurisdictions as the locations of the headquarters. Germany (DE) and France (FR) also have many headquarters, compared with the other jurisdictions. Second bar plot shows regarding the ``holding" companies. The Netherlands (NL), the US, and the UK (GB) are remarkable as the locations of the ``holding" companies. Germany (DE) and Spain (ES) have more the ``holding" companies, compared with the other jurisdictions. Third bar plot shows regarding the ``holding and conduit" companies. The US, the UK (GB), and the Netherlands (NL) are still noticeable as the locations of the ``holding and conduit" companies. Germany (DE) and Spain (ES) have more the ``holding and conduit" companies, compared with the other jurisdictions. Fourth bar plot shows regarding the ``conduit" companies. The distribution of the ``conduit" companies is slightly different from those of the ``holding" companies and the ``holding and conduit" companies. In addition to the Netherlands (NL) and the UK (GB),  has many ``conduit" companies are located in the Netherlands (NL), Hong Kong (HK), the UK (GB), and Germany (DE), the British Virgin Islands (VG) is also perceptible. Many key companies are located in Europe, especially the Netherlands and the UK, although the two jurisdictions are not at the top five jurisdictions as the locations of the headquarters. The two jurisdictions are empirically known for their taxation merits, which MNCs can exploit through the acquisition of intermediate companies for the international taxation (Eicke 2009). Meanwhile, China, France, Austria, Italy, and Japan are not at the top five jurisdictions of the number of the key companies, even though these jurisdictions have many affiliates (see Figure \ref{affiliates_jurisdiction}). \par
On the other hand, the right bar plots of Figure \ref{distribution} shows the distributions of the headquarters and the key companies in the bow-tie structure of the GWCC of the GON. First bar plot shows regarding the headquarters. Even though the bow-tie structure (the GWSS) accounts for only 11.433\% of the entire GON, most of the headquarters are located in the GSCC, the IN component, or the TE. Moreover, it is noticeable that their key companies are mostly located in the IN component of the bow-tie structure, although the IN component accounts for only 1.95\% of the entire GON.

\subsection*{Withholding Tax}
We examine the relationship between the number of the key companies and the possibility of being used for treaty shopping in detail because the major incentives for establishing key companies is to avoid the withholding tax imposed on the profits of MNCs. To examine whether a relationship exists between the location of the key companies and the possibility to be used for treaty shopping, we use the withholding tax centrality as an indicator of the possibility to be used for treaty shopping (Nakamoto and Ikeda 2018). The withholding tax centrality is calculated based on the withholding tax rates imposed by each jurisdiction on dividends. The higher the withholding tax centrality is, the more a jurisdiction is likely to be used for treaty shopping.\par

  \begin{figure}[h]
\centering
  \includegraphics[width=6.0cm, angle=270]{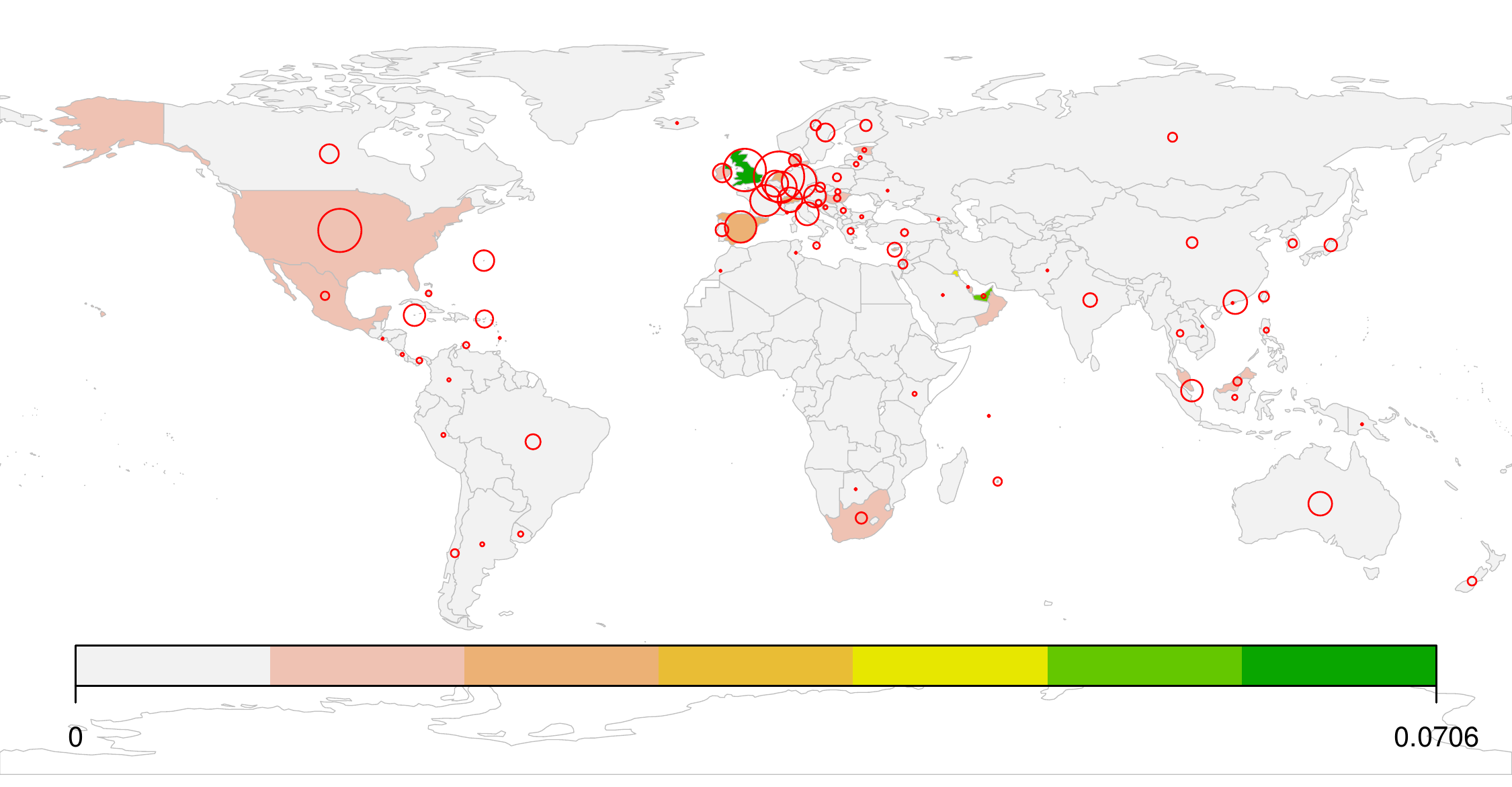}
  \caption{{ \bf ``Holding" companies and Treaty shopping.}
The size of the circles indicates the number of ``holding" companies and the depth of the color indicates the possibility to be used for treaty shopping by the jurisdiction.}
  \label{fig:base-holding}
  \end{figure}

  \begin{figure}[h]
\centering
  \includegraphics[width=6.0cm, angle=270]{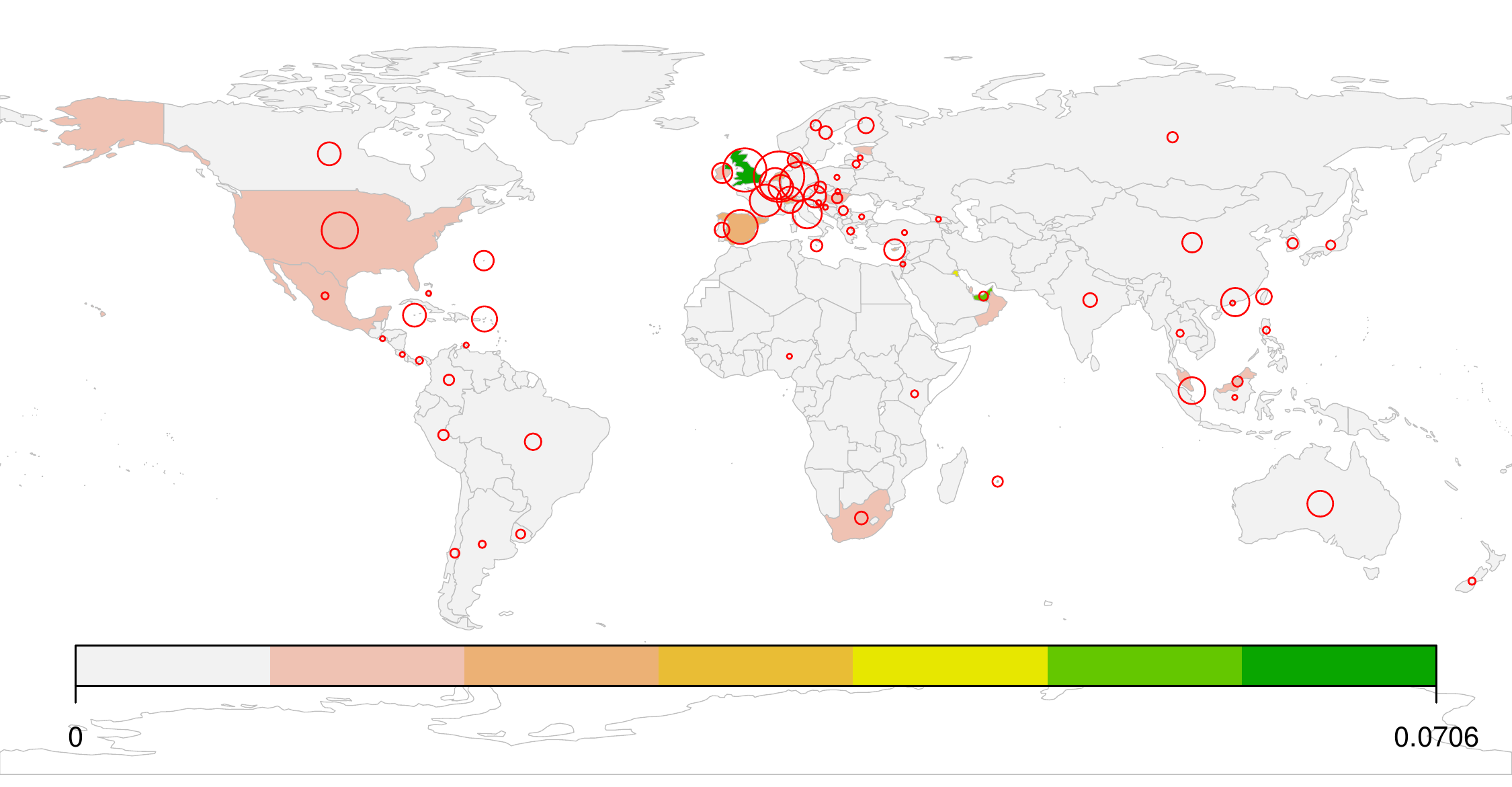}
  \caption{{\bf ``Holding \& conduit" companies and Treaty shopping.}
The size of the circles indicates the number of ``holding and conduit" companies and the depth of the color indicates the possibility to be used for treaty shopping by the jurisdiction.}
  \label{fig:base-and-conduit-firms-on-maps}
  \end{figure}

  \begin{figure}[h]
\centering
  \includegraphics[width=6.0cm, angle=270]{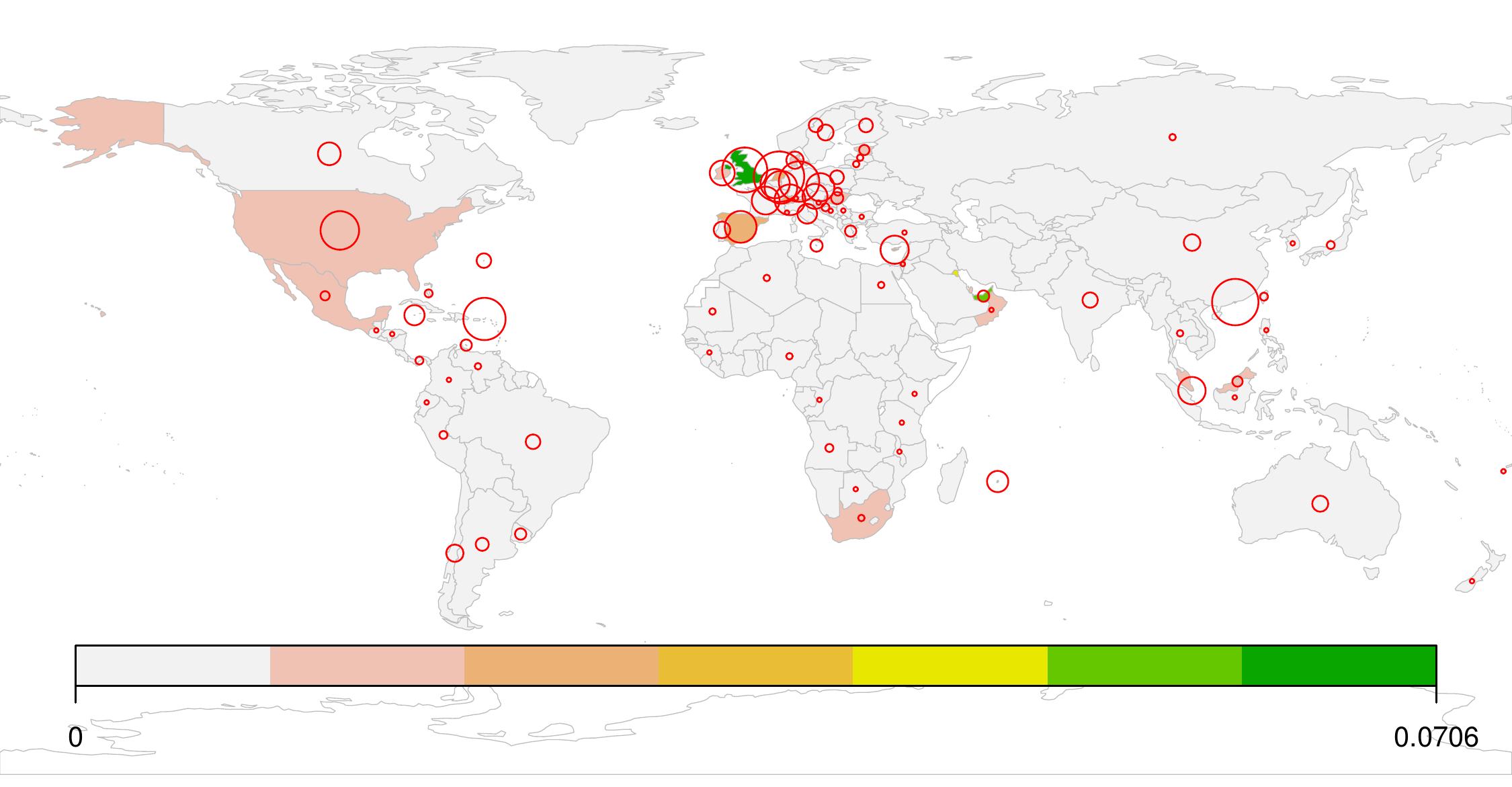}
  \caption{{\bf ``Conduit" companies and Treaty shopping.}
The size of the circles indicates the number of ``conduit" companies and the depth of the color indicates the possibility to be used for treaty shopping by the jurisdiction.}
  \label{fig:conduit-firms-on-maps}
  \end{figure}

In Figures \ref{fig:base-holding}-\ref{fig:conduit-firms-on-maps}, the size of the circles shows the geographic distributions of the key companies used for treaty shopping. The depth of the color shows the withholding tax centrality, which represents the possibility to be used for treaty shopping. The more jurisdictions that are likely to be used for treaty shopping, the darker the color. There seems a certain relationship between the number of the key companies and the possibility of being used for treaty shopping.\par

\begin{figure}[ht]
\centering
\subfigure[Holding companies]{
\includegraphics[width=3.7cm, angle=270]{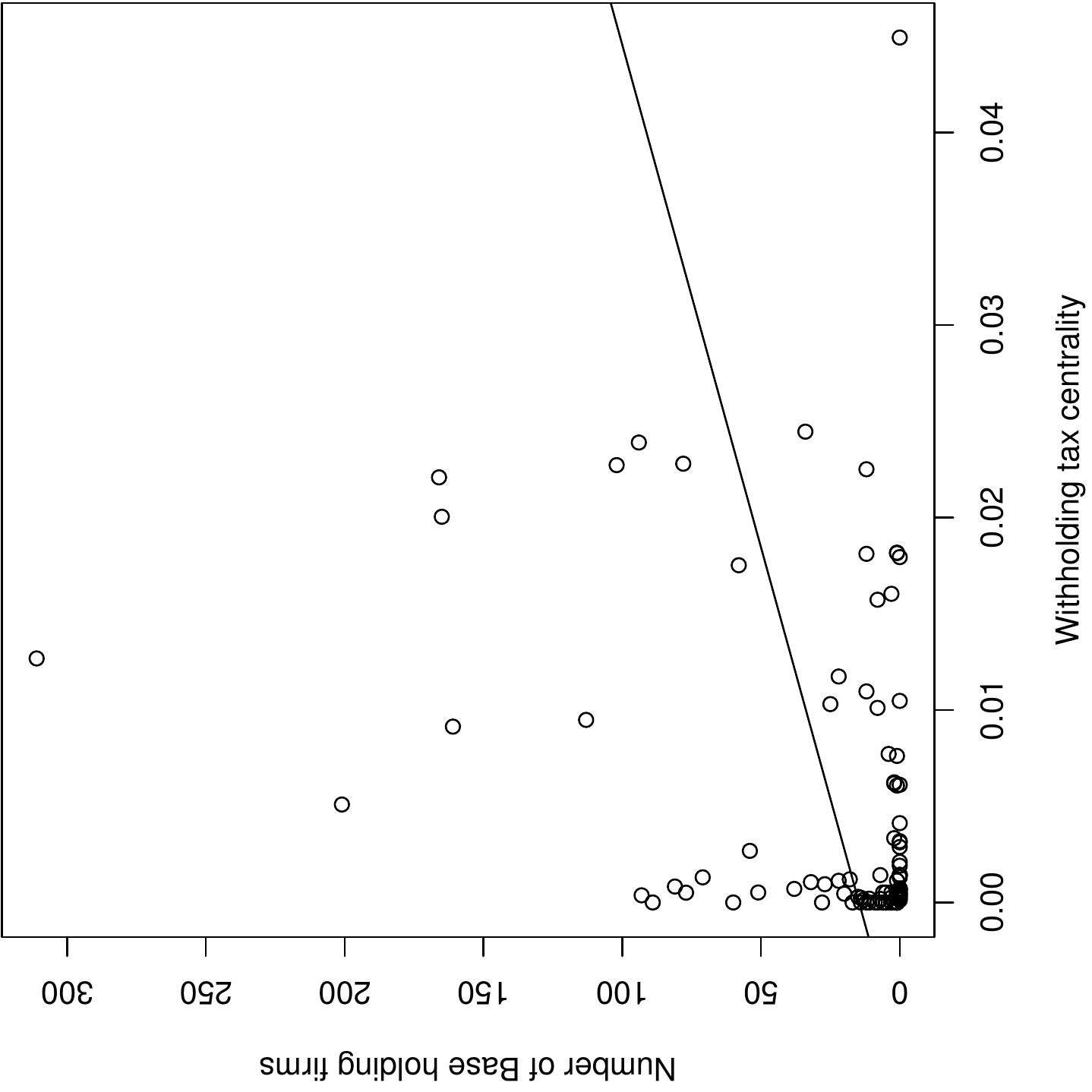}
}
\subfigure[Holding \& conduit companies]{
\includegraphics[width=3.7cm, angle=270]{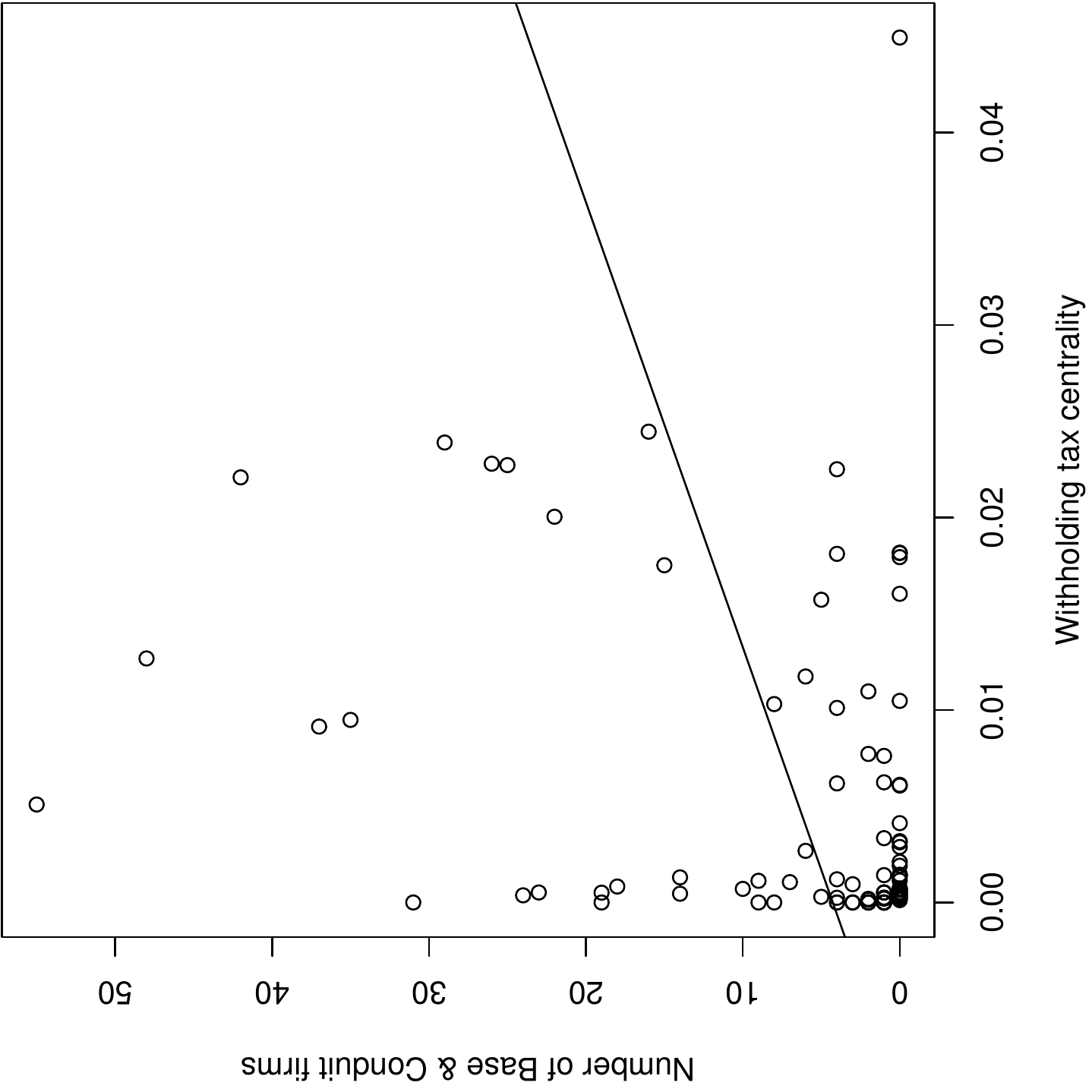}
}
\subfigure[Conduit companies]{
\includegraphics[width=3.7cm, angle=270]{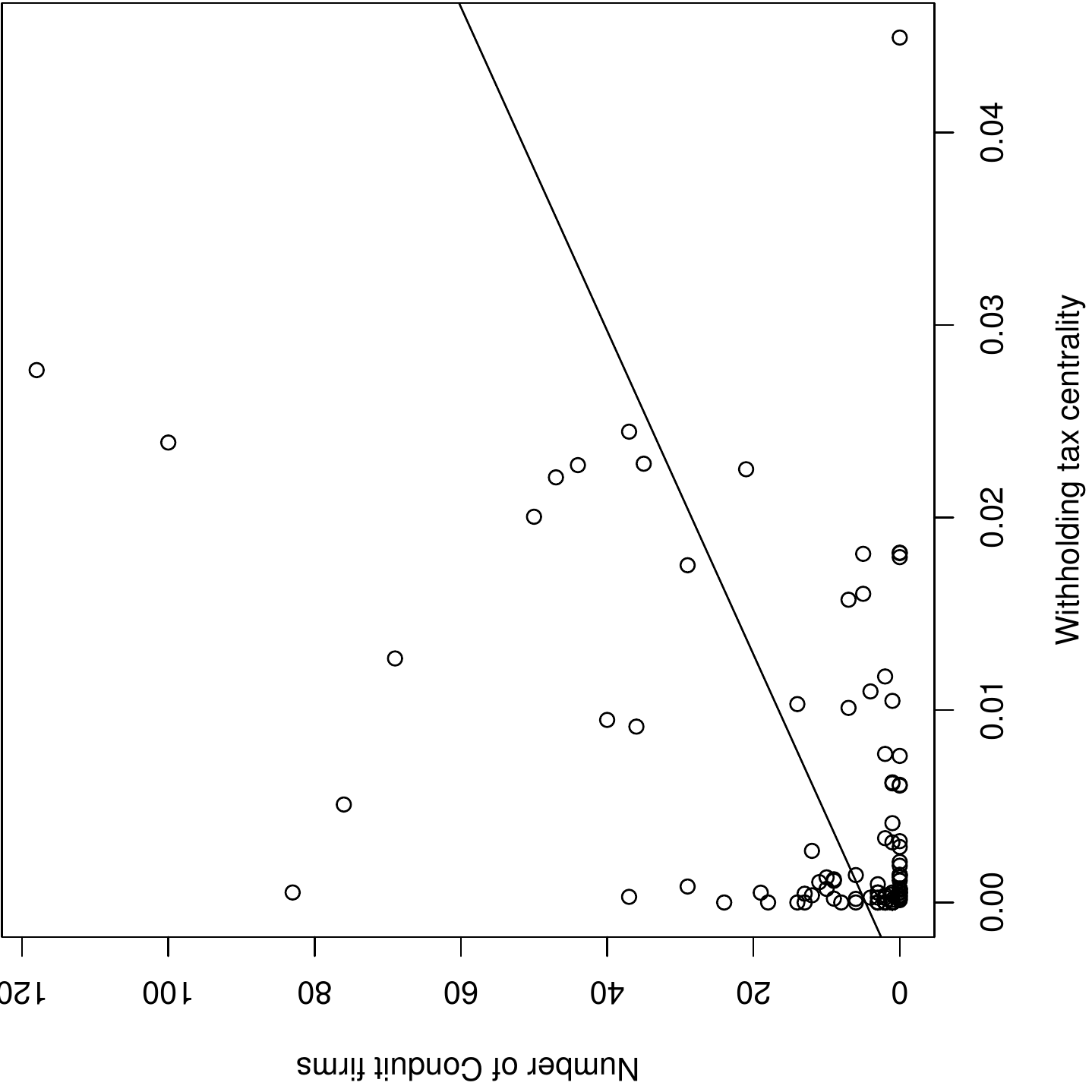}
}
\caption{{\bf Key companies and withholding tax centrality.} The vertical axis indicates a number of key companies and the horizontal axis is the withholding tax centrality, which indicates the possibility to be used for treaty shopping.}
\label{fig:plot}       
\end{figure}

To examine the relationship between the location of the key companies and the possibility to be used for treaty shopping more precisely, we perform regression analyses using the number of the key companies as a target variable and the possibility to be used for treaty shopping as an explanatory variable. Figure \ref{fig:plot} shows the results of the regression analyses. Regarding ``holding" companies, an obtained regression equation is $y = 14.671 + 1914.126 x$ and the adjusted R-squared is 0.0888. The t-value of the intercept is 2.856, and the its p-value is 0.005154; this is significant at 1\%. The t-value of the withholding tax centrality is 3.395, and its p-value was 0.000964; this is significant at 0.1\%. Regarding ``holding and conduit" companies, an obtained regression equation is $y = 4.267 + 432.241 x$ and adjusted R-squared is 0.08672. The t-value of the intercept is 3.475; and its p-value is 0.000751, and it is significant at 0.1\%. The t-value of the withholding tax centrality is 3.283; and its p-value is 0.001406; this is significant at 1\%. Regarding ``conduit" companies, an obtained regression equation is $y = 4.678 + 1188.987 x$ and the adjusted R-squared is 0.2073. The t-value of the intercept is 2.365; and its p-value is 0.0197; this is significant at 10\%. The t-value of the withholding tax centrality is 5.551; its p-value is $1.9\times10^{-7}$; this is significant at 0.1\%. Figure \ref{fig:plot} is scattered plots with the number of key companies identified on the X-axis and the withholding tax centrality on the Y-axis. The straight line is the obtained regression line.\par
The locations of the key companies clearly correlate with the possibilities for the locations to be used for treaty shopping. It is found that treaty shopping has a greater influence on the number of ``holding" companies and ``conduit" companies because their coefficient of withholding tax centrality is high. It is also worthy of attention that the withholding tax centrality can explain the number of ``conduit" companies by about 20\%. 

\subsection*{Ownership Chains}
Establishment of intermediate companies or the key companies leads to a diversion of FDI. We analyze the investments via the key companies. Tables \ref{tab:base-chain}-\ref{tab:conduit-chain} show for what jurisdictions the top three jurisdictions with many key companies are intermediate destinations. The column of ``Subsidiary" shows the location of affiliates where the key companies hold shares, and the column of ``Shareholder" shows the location of affiliates holding shares of the key companies.

\begin{table}[h]
\centering
 \caption{Direct shareholders and subsidiaries of the ``holding" companies.}
 \begin{tabular}{cc cc cc cc cc cc}
\toprule
 \multicolumn{4}{c}{the Netherlands} & \multicolumn{4}{c}{the United States} & \multicolumn{4}{c}{the United Kingdom}\\
\cmidrule(lr){1-4} \cmidrule(lr){5-8} \cmidrule(lr){9-12}
 \multicolumn{2}{c}{Subsidiary} & \multicolumn{2}{c}{Shareholder} & \multicolumn{2}{c}{Subsidiary} &  \multicolumn{2}{c}{Shareholder} & \multicolumn{2}{c}{Subsidiary} & \multicolumn{2}{c}{Shareholder}\\
\cmidrule(lr){1-4} \cmidrule(lr){5-8} \cmidrule(lr){9-12}
 IE &  9\% & US & 35\% & CA & 18\% & GB & 14\% & US & 15\% & US & 34\%\\
 US &  8\% & GB & 11\% & GB & 13\% & CA & 12\% & ZA &  5\% & n.a. & 16\%\\
 n.a. &  7\% & FR & 10\% & IE &  5\% & JP & 12\% & IN &  5\% & FR &  7\% \\
 DE &  7\% & DE &  6\% & JP & 4\% & DE & 10\% & NL &  4\% & CH &  4\% \\
 FR &  6\% & CH & 4\% & MX & 4\% & FR &  8\% & IE &  3\% & DE &  4\% \\
\bottomrule \noalign{\smallskip}
\end{tabular}
\begin{flushleft}
IE is Ireland; DE is Germany; FR is France; GB is the UK; CH is Switzerland; CA is Canada; JP is Japan; MX is Mexico; ZA is South Africa; IN is India; NL is the Netherlands; and n.a. is no information.
\end{flushleft}
 \label{tab:base-chain}
 \end{table}

Table \ref{tab:base-chain} indicates the ``holding" companies located in the Netherlands, the US, and the UK, which are the top three jurisdictions with many ``holding" companies. The investments through the Netherlands (NL) are the major investments among Western jurisdictions, such as Ireland (IE) and Germany (DE). The investments through ``holding" companies located in the United States are the investments relating to Canada (CA) and Mexico (MX), which are geographically close to the US. It is also the investments related to Japan (JP), which is an East Asian jurisdiction. The investments through the United Kingdom (GB) are the investments by the Commonwealth jurisdictions, such as South Africa (ZA) and India (IN).

\begin{table}[h]
\centering
\caption{Direct shareholders and subsidiaries of the ``holding and conduit" companies.}
\begin{tabular}{cc cc cc cc cc cc}
\toprule
\multicolumn{4}{c}{the United States} & \multicolumn{4}{c}{the United Kingdom} & \multicolumn{4}{c}{the Netherlands}\\
\cmidrule(lr){1-4} \cmidrule(lr){5-8} \cmidrule(lr){9-12}
\multicolumn{2}{c}{Subsidiary} & \multicolumn{2}{c}{Shareholder} & \multicolumn{2}{c}{Subsidiary} &  \multicolumn{2}{c}{Shareholder} & \multicolumn{2}{c}{Subsidiary} & \multicolumn{2}{c}{Shareholder}\\
\cmidrule(lr){1-4} \cmidrule(lr){5-8} \cmidrule(lr){9-12}
CA & 20\% & GB & 15\% & ZA & 31\% & US & 29\% & UA & 17\% & US & 19\%\\
CN & 13\% & FR & 15\% & AU &  9\% & n.a. & 27\% & BE &  9\% & GB & 14\%\\
LU &  9\% &	CH & 11\% & LU &  7\% & ZA & 12\% & GB &  6\% & PL &  8\% \\
GB &  9\% & DE & 11\% & BR &  5\% & FR &  5\% & NG &  5\% & FR &  7\% \\
HK &  5\% & CA &  7\% & VG &  5\% & BM &  4\% & n.a. &  4\% & DE &  7\% \\
\bottomrule \noalign{\smallskip}
 \end{tabular}
\begin{flushleft}
CA is Canada; CN is mainland China; LU is Luxembourg; GB is the UK; HK is Hong Kong; FR is France; CH is Switzerland; DE is Germany; ZA is South Africa; AU is Austria; BR is Brazil; VG is the British Virgin Islands; BM is the Bermuda Islands; UA is the United Arab Emirates; BE is Belgium; NG is Nigeria; PL is Poland; and n.a. is no information.
\end{flushleft}
 \label{tab:regional-chain}
 \end{table}

Table \ref{tab:regional-chain} provides details of the ``holding and conduit" companies located in the US, the UK, and the Netherlands, which are the top three jurisdictions with many ``holding and conduit" companies. The investments through the US include many investments related to China (CN), which is an East Asian jurisdiction. The investments through the UK (GB) consist of many investments from the Bermuda Islands (BM) and the British Virgin Islands (VG), which are British overseas territories and known as so-called tax havens. In addition, ``holding and conduit" companies located in the UK make much investment in South Africa (ZA). It is noteworthy that South Africa (ZA) occupies a high proportion of both columns of ``Subsidiary" and ``Shareholder." The Investments in South Africa (ZA) from South Africa (ZA) via the UK (GB) suggests the existence of round tripping (IMF 2004). The ``holding and conduit" companies located in the Netherlands (NL) make much investment in the United Arab Emirates (UA) due to oil companies. 

 \begin{table}[h]
 \centering
 \caption{Direct shareholders and subsidiaries of the ``conduit" companies.}
 \begin{tabular}{cc cc cc cc cc cc}
\toprule
 \multicolumn{4}{c}{the Netherlands} & \multicolumn{4}{c}{Hong Kong} & \multicolumn{4}{c}{the United Kingdom}\\
\cmidrule(lr){1-4} \cmidrule(lr){5-8} \cmidrule(lr){9-12}
 \multicolumn{2}{c}{Subsidiary} & \multicolumn{2}{c}{Shareholder} & \multicolumn{2}{c}{Subsidiary} &  \multicolumn{2}{c}{Shareholder} & \multicolumn{2}{c}{Subsidiary} & \multicolumn{2}{c}{Shareholder}\\
\cmidrule(lr){1-4} \cmidrule(lr){5-8} \cmidrule(lr){9-12}
 DE & 13\% & ES & 14\% & CN & 60\% & KY & 24\% & ES & 28\% & NL & 12\%\\
 GB	& 10\% & GB & 14\% & VG & 11\% & BM & 13\% & IE & 10\% & DE & 10\%\\
 PL	&  7\% & US & 13\% & TW &  6\% & VG & 11\% & ID &  8\% & IE &  9\% \\
 BE	&  7\% & DE & 11\% & NL &  4\% & FR &  6\% & DE &  7\% & US & 7\% \\
 ES	&  7\% & CH &  8\% & BM &  2\% & CN &  5\% & NL &  6\% & AU &  7\% \\
\bottomrule \noalign{\smallskip}
 \end{tabular}
\begin{flushleft}
DE is Germany; GB is the UK; PL is Poland; BE is Belgium; ES is Spain; CH is Switzerland; CN is mainland China; VG is the British Virgin Islands; TW is Taiwan; NL is the Netherlands; BM is the Bermuda Islands; FR is France; IE is Ireland; ID is Indonesia; and AU is Austria.
\end{flushleft}
 \label{tab:conduit-chain}
 \end{table}

Table \ref{tab:conduit-chain} shows the results of ``conduit" companies located in the Netherlands, Hong Kong, and the UK, which are the top three jurisdictions with many ``conduit" companies. The investments through the ``conduit" companies located in the Netherlands (NL) are mainly the investments related to Western jurisdictions. The investments through Hong Kong (HK) are mainly investments in mainland China (CN) and Taiwan (TW). In addition, the interesting point is that both columns ``Shareholder" and ``Subsidiary" have the Cayman Islands (KY), the British Virgin Islands, and the Bermuda Islands for investments. These locations are British Overseas Territories used as tax havens. The investments through the ``conduit" companies located in the UK (GB) are many investments related to Indonesia (ID) and Australia (AU).\par
As a general trend, the key companies are close to the geographical jurisdictions of the headquarters. In addition to the trend, not a few investments through the US are related to East Asian jurisdictions such as Japan (JP) and China (CN). The investments through the UK or former leasehold land Hong Kong are mainly the investments in the Commonwealth and the British overseas territories. It is notable that affiliates located in so-called tax havens are found to have connections with the ``holding and conduit" or ``conduit" companies. This implies affiliates related to tax havens are the lower ownership layer in the ownership structures.

\subsection*{Headquarters}
Finally, we analyze whether there is a difference between the key companies in the location of the headquarters. In Table \ref{tab:ultimate parent}, we use our model to show the location of the headquarters holding the key companies. In all the key companies, which are ``holding," ``holding and conduit," and ``conduit" companies, the headquarters located in the US own the most key companies. The reason is that approximately 25\% of the MNCs listed in the Fortune Global 500 (subject to this analysis) are located in the US. In contrast, the MNCs in the UK own many key companies as compared with the number of the headquarters of the MNCc listed in the Fortune Global 500 (subject to this analysis). It is suggested that MNCs in the UK tend to have many affiliates that can play an important role in international profit shifting as compared with MNCs in other jurisdictions.\par

 \begin{table}[h]
\centering
 \caption{Top five locations of the headquarters holding the key companies.}
 \begin{tabular}{cc cc cc}
\toprule
 \multicolumn{2}{c}{Holding} & \multicolumn{2}{c}{Holding \& Conduit}  & \multicolumn{2}{c}{Conduit} \\
\cmidrule(lr){1-2} \cmidrule(lr){3-4} \cmidrule(lr){5-6}
 US & 27\% & US & 30\% & US & 27\% \\
 GB & 13\% & GB & 19\% & GB & 16\% \\
 JP & 11\% & CN & 9\% & CN & 13\% \\
 DE & 11\% & DE & 8\% & DE & 10\% \\
 CN &  8\% & JP & 7\% & JP &  8\% \\ 
\bottomrule \noalign{\smallskip}
 \end{tabular}
\begin{flushleft}
GB is the UK; JP is Japan; DE is Germany; and CN is mainland China.
\end{flushleft}
 \label{tab:ultimate parent}
 \end{table}

 \begin{table}[h]
 \centering
 \caption{``Holding" companies of top three jurisdictions.}
 \begin{tabular}{cc cc cc}
\toprule
 \multicolumn{2}{c}{the United States} & \multicolumn{2}{c}{the United Kingdom} & \multicolumn{2}{c}{Japan}\\
\cmidrule(lr){1-2} \cmidrule(lr){3-4} \cmidrule(lr){5-6}
 GB	& 15\% & NL & 15\% & US & 22\% \\
 NL	& 13\% & US & 12\% & GB & 13\% \\
 DE	&  9\% &	ES &  7\% & NL & 12\% \\
 LU	&  8\% & LU &  7\% & DE & 6\% \\
 FR	&  5\% & DE & 6\% & SG & 5\% \\ 
\bottomrule \noalign{\smallskip}
 \end{tabular}
\begin{flushleft}
GB is the UK; NL is the Netherlands; DE is Germany; LU is Luxembourg; FR is France; ES is Spain; and SG is Singapore.
\end{flushleft}
 \label{tab:base-top3}
 \end{table}

Tables \ref{tab:base-top3}-\ref{tab:conduit-top3} summarize the locations of the key companies held by MNCs in the top three jurisdiction in Table \ref{tab:ultimate parent}. Moreover, the tables show whether there is any difference in the location of the key companies and the locations of MNCs. Table \ref{tab:base-top3} shows ``holding" companies are mainly located in Western jurisdictions for any MNCs in the top three jurisdictions. MNCs in Japan (JP) also own many ``holding" companies in Singapore (SG). This seems to reflect that many Japanese MNCs prefer to establish their affiliates in Singapore (SG) when doing business in Southeast Asia.

 \begin{table}[h]
 \centering
 \caption{``Holding and conduit" companies of top three jurisdictions.}
 \begin{tabular}{cc cc cc}
\toprule
 \multicolumn{2}{c}{the United States} & \multicolumn{2}{c}{the United Kingdom} & \multicolumn{2}{c}{China}\\
\cmidrule(lr){1-2} \cmidrule(lr){3-4} \cmidrule(lr){5-6}
NL	& 12\% & GB & 11\% & HK & 12\% \\
 US	& 12\% & NL &  9\% & CN &  9\% \\
 GB	&  9\% &	DE &  9\% & VG &  9\% \\
 DE	&  7\% & US &  8\% & KY &  8\% \\
 ES	&  7\% & ES &  6\% & US &  8\% \\ 
\bottomrule \noalign{\smallskip}
 \end{tabular}
\begin{flushleft}
NL is the Netherlands; GB is the UK; DE is Germany; ES is Spain; HK is Hong Kong; CN is mainland China; VG is the British Virgin Islands; and KY is the Cayman Islands.
\end{flushleft}
 \label{tab:regional-top3}
 \end{table}

 \begin{table}[h]
 \centering
 \caption{``Conduit" companies of top three jurisdictions.}
 \begin{tabular}{cc cc cc}
\toprule
 \multicolumn{2}{c}{the United States} & \multicolumn{2}{c}{the United Kingdom} & \multicolumn{2}{c}{China}\\
\cmidrule(lr){1-2} \cmidrule(lr){3-4} \cmidrule(lr){5-6} 
GB & 17\% & GB & 17\% & CN & 29\% \\
 ES & 14\% & IE & 10\% & HK & 24\% \\
 DE &  9\% &	US & 10\% & VG &  9\% \\
 US	&  8\% & ES &  9\% & KY &  6\% \\
 RU	&  6\% & DE & 6\% & BM &  6\% \\ 
\bottomrule \noalign{\smallskip}
 \end{tabular}
\begin{flushleft}
GB is the UK; ES is Spain; DE is Germany; RU is Russia; IE is Ireland; CN is mainland China; HK is Hong Kong; VG is the British Virgin Islands; KY is the Cayman Islands; and the BM is the Bermuda Islands.
\end{flushleft}
 \label{tab:conduit-top3}
 \end{table}

On the other hand, Tables \ref{tab:regional-top3}-\ref{tab:conduit-top3} show ``holding and conduit" and ``conduit" companies, which are at the lower layer of the ownership structure, are located in the jurisdictions where their headquarters are located for any MNCs located in the top three jurisdictions. In the case of the US MNCs, 12\% of ``holding and conduit" companies are located in the US, in the case of the UK MNCs, 11\% of  ``holding and conduit" companies are located in the UK, and in the case of Chinese MNCs, 9\% of ``holding and conduit" companies are located in mainland China. Similarly, 8\% of ``conduit" companies the US MNCs have are located in the US, 17\% of ``conduit" companies the UK MNCs have are located in the UK, and 29\% of ``conduit" companies Chinese MNCs have are located in mainland China. Because the key companies satisfies the third country type (see the Key Companies subsections in the Model section), affiliates located in different jurisdictions from the location of the headquarter possess key companies located in the jurisdiction where their headquarters are located. This suggests the possibility of performing round tripping (IMF 2004). In addition, it is noteworthy that Chinese MNCs have many affiliates in jurisdictions known as so-called tax havens, such as the Cayman Islands (KY), the British Virgin Islands (VG), and the Bermuda Islands (BM).

\section*{Conclusions}
In the global economy, the existence of intermediate company is becoming a problem. The purpose of this paper is to identify intermediate companies, or key companies, that are at high risk in international profit shifting in order to clarify the actual situation of the intermediate companies. We analyzed large MNCs since large MNCs have a high risk of international profit shifting.\par
Our proposed model focuses on the position the affiliates occupy in the ownership structure of each MNC, rather than using financial information as used by previous studies. In analyzing the ownership structure, the GON was constructed using the information in the Orbis database to establish the relationship between the key companies and MNCs. Thus, we analyzed the basic features of the GON used in this analysis. We confirmed the validity of our model to identify that affiliates play an important role in international profit shifting of five selected MNCs. We use MNCs listed in the Fortune Global 500 (subject to this analysis) for our model. It was found that many identified key companies exist in well-known jurisdictions such as the Netherlands and the UK. On the GON, it was found that such identified companies concentrated in the IN component of the bow-tie structure.\par
Considering the results in the context of international taxation, we found that there is a relationship between the number of identified key companies and treaty shopping. Thus, it is clear that there is a difference in the investment's routes depending on the location of key companies. For example, the investments related to East Asian jurisdictions often pass through the key companies located in the US and the investments related to the Commonwealth tend to pass through the key companies located in the UK. In addition, it was revealed that MNCs in the UK held the relatively large number of key companies, and Chinese MNCs held many ``conduit" companies in the so-called tax havens. It is suggested that key companies not only receive a strong influence of the withholding tax rate but also strongly reflect the historical, economic, and political relationships of the location of key companies. The key companies are a microcosm of such various relationships. The future work should focus on network properties of the ownership structures in more detail and cooperative analysis in the cross-section to better understand the GON.

\section*{Competing interests}
The authors declare that they have no competing interests.

\section*{Funding}
The present study was supported partly by the Ministry of Education, Science, Sports, and Culture, Grants-in-Aid for Scientific Research (B), Grant No. 17KT0034 (2017-2019), and by MEXT as Exploratory Challenge s on Post-K computer (Studies of Multi-level Spatiotemporal Simulation of Socioeconomic Phenomena).

\section*{Author's contributions}
All co-authors reviewed, edited the study, and read the final text of this study carefully. TN provided key contributions in implementing the network analysis and performed the visualization. TN also made a survey of the related work, conceptualized the problem, and wrote the draft. AC performed bow-tie structure analysis and community identification of the network. YI supervised the project and contributed formulation of the centralities and development of the hierarchical identification algorithm.

\section*{Acknowledgments}
The authors wish to acknowledge Tadao Okamura and Hiroaki Takashima for valuable advice. The authors would like to thank Enago (www.enago.jp) for the English language review,


\end{document}